\documentstyle[11pt]{article}
\begin{document}
\newcommand{\Si}{\Sigma}
\newcommand{\tr}{{\rm tr}}
\newcommand{\ad}{{\rm ad}}
\newcommand{\Ad}{{\rm Ad}}
\newcommand{\ti}[1]{\tilde{#1}}
\newcommand{\om}{\omega}
\newcommand{\Om}{\Omega}
\newcommand{\de}{\delta}
\newcommand{\al}{\alpha}
\newcommand{\te}{\theta}
\newcommand{\vth}{\vartheta}
\newcommand{\be}{\beta}
\newcommand{\la}{\lambda}
\newcommand{\La}{\Lambda}
\newcommand{\D}{\Delta}
\newcommand{\ve}{\varepsilon}
\newcommand{\ep}{\epsilon}
\newcommand{\vf}{\varphi}
\newcommand{\G}{\Gamma}
\newcommand{\ka}{\kappa}
\newcommand{\ip}{\hat{\upsilon}}
\newcommand{\Ip}{\hat{\Upsilon}}
\newcommand{\ga}{\gamma}
\newcommand{\ze}{\zeta}
\newcommand{\si}{\sigma}
\def\bfa{{\bf a}}
\def\bfb{{\bf b}}
\def\bfc{{\bf c}}
\def\bfd{{\bf d}}
\def\bfm{{\bf m}}
\def\bfn{{\bf n}}
\def\bfp{{\bf p}}
\def\bfu{{\bf u}}
\def\bfv{{\bf v}}
\def\bft{{\bf t}}
\def\bfx{{\bf x}}
\def\bfg{{\bf g}}
\newcommand{\li}{\lim_{n\rightarrow \infty}}
\newcommand{\mat}[4]{\left(\begin{array}{cc}{#1}&{#2}\\{#3}&{#4}
\end{array}\right)}
\newcommand{\thmat}[9]{\left(
\begin{array}{ccc}{#1}&{#2}&{#3}\\{#4}&{#5}&{#6}\\
{#7}&{#8}&{#9}
\end{array}\right)}
\newcommand{\beq}[1]{\begin{equation}\label{#1}}
\newcommand{\eq}{\end{equation}}
\newcommand{\beqn}[1]{\begin{eqnarray}\label{#1}}
\newcommand{\eqn}{\end{eqnarray}}
\newcommand{\p}{\partial}
\newcommand{\di}{{\rm diag}}
\newcommand{\oh}{\frac{1}{2}}
\newcommand{\su}{{\bf su_2}}
\newcommand{\uo}{{\bf u_1}}
\newcommand{\GL}{{\rm GL}(N,{\bf C})}
\newcommand{\SL}{{\rm SL}(3,{\bf C})}
\def\sln{{\rm sl}(N,{\bf C})}
\newcommand{\gl}{gl(N,{\bf C})}
\newcommand{\PSL}{{\rm PSL}_2({\bf Z})}
\def\f1#1{\frac{1}{#1}}
\def\lb{\lfloor}
\def\rb{\rfloor}
\newcommand{\rar}{\rightarrow}
\newcommand{\upar}{\uparrow}
\newcommand{\sm}{\setminus}
\newcommand{\ms}{\mapsto}
\newcommand{\bp}{\bar{\partial}}
\newcommand{\bz}{\bar{z}}
\newcommand{\bA}{\bar{A}}
\newcommand{\sect}[1]{\setcounter{equation}{0}\section{#1}}
\renewcommand{\theequation}{\thesection.\arabic{equation}}
\newtheorem{predl}{Proposition}[section]
\newtheorem{defi}{Definition}[section]
\newtheorem{rem}{Remark}[section]
\newtheorem{cor}{Corollary}[section]
\newtheorem{lem}{Lemma}[section]
\newtheorem{theor}{Theorem}[section]

\vspace{0.3in}
\begin{flushright}
 ITEP-TH-15/00\\
\end{flushright}
\vspace{10mm}
\begin{center}
{\Large\bf
Hamiltonian Algebroid Symmetries\\
in $W$-gravity and Poisson sigma-model}\\
\vspace{5mm}
A.M.Levin\\
{\sf Institute of Oceanology, Moscow, Russia,} \\
{\em e-mail andrl@landau.ac.ru}\\
M.A.Olshanetsky
\\
{\sf Institute of Theoretical and Experimental Physics, Moscow, Russia,}\\
{\em e-mail olshanet@heron.itep.ru}\\

\vspace{5mm}
\end{center}
\begin{abstract}
Starting from a Lie algebroid ${\cal A}$ over a space $V$ we  lift
its action to the canonical transformations on the principle affine bundle ${\cal R}$
over the cotangent bundle $T^*V$. Such lifts are classified by the first cohomology
$H^1({\cal A})$. The resulting object is the Hamiltonian algebroid ${\cal A}^H$ over
${\cal R}$ with the anchor map from $\G({\cal A}^H)$ to Hamiltonians of canonical
transformations. Hamiltonian algebroids generalize the Lie algebras of
canonical transformations. We prove that the BRST operator for ${\cal A}^H$
is cubic in the ghost fields as in the Lie algebra case. To illustrate this construction
we analyze two topological field theories. First, we define a Lie algebroid over
the space $V_3$ of $\SL$-opers  on a Riemann curve $\Si_{g,n}$ of genus $g$
with $n$ marked points. The sections of this algebroid are the second order
differential operators on $\Si_{g,n}$. The algebroid is lifted to the Hamiltonian
algebroid over the phase space of $W_3$-gravity. We describe the BRST operator leading
to the moduli space of $W_3$-gravity. In accordance with the general construction
the BRST operator is cubic in the ghost fields. We present the Chern-Simons explanation
of our results. The second example is the Hamiltonian algebroid structure in the Poisson
sigma-model invoked by Cattaneo and Felder to describe the Kontsevich deformation
quantization formula. The hamiltonian description of the Poisson sigma-model leads
to the Lie algebraic form of the BRST operator.
\end{abstract}

\vspace{0.15in}
\bigskip
\title
\maketitle
\date{}

\section {Introduction}
\setcounter{equation}{0}
 Lie groups by no means exhaust the symmetries in gauge theories.
Their importance is related to the natural geometric structures
defined by a group action in accordance with the Erlanger
program of F.Klein. In fact, the first class constraints in Hamiltonian systems
generate the canonical transformations of the phase space which generalize the
Lie group actions \cite{HT}.  Our main interest lies in topological field
theories,
where the factorization with respect to the canonical gauge transformations
may lead to generalized  deformations of corresponding moduli spaces.

There exists a powerful method to treat such
types of structures. It is the BRST method that is applicable in
Hamiltonian and Lagrangian forms \cite{BFV}. The BRST operator
corresponding to arbitrary first class constraints acquires the most general
form. An intermediate step in this direction is the canonical transformations
generated by the quasigroups \cite{Ba,KM}.
The BRST operator for the quasigroup action has the same form as for
the Lie group case.

Here we consider the quasigroup symmetries that
constructed by means of special kind transformations of the "coordinate
space" $V$. These transformations along with the coordinate space $V$ are
the Lie groupoids, or their infinitesimal version - the Lie algebroids
${\cal A}$ \cite{Ma,We}.
We lift them to the cotangent bundle $T^*V$, or, more generally, on
 principle homogeneous space ${\cal R}$ over
the cotangent bundle $T^*V$ in a such way that they become
the quasigroup transformations. The infinitesimal form of them we call
the Hamiltonian algebroid ${\cal A}^H$ related to the Lie algebroid ${\cal A}$.
The Hamiltonian algebroid is analog of the Lie algebra of symplectic
vector fields with respect to the canonical symplectic structure on
${\cal R}$ or $T^*V$.
These lifts are classified by the first cohomology group $H^1({\cal A})$.
To put it otherwise, the first class constraints in this case produce the
quasigroup symmetries.  As a result the BRST operator has the same structure as
for the Lie algebras transformations.

We present two examples of topological field theories
with these symmetries. The first example is the
$W_3$-gravity \cite{P,BFK,GLM} and related to this theory
 generalized deformations
of complex structures of Riemann curves by the second order differential
operators. This theory is a generalization
of $2+1$-gravity ($W_2$-gravity) \cite{Ca}, where the space component has a topology of
a Riemann curve of genus $g$ with $n$  marked points $\Si_{g,n}$.
The Lie algebra symmetries in $W_2$-gravity is the algebra of
smooth vector fields on $\Si_{g,n}$. After killing the gauge degrees of freedom
one comes  to the moduli space of projective
 structures on $\Si_{g,n}$. These structures can be described by the BRST method
which is straightforward in this case. The case of $W_N$-gravity $(N>2)$ is more subtle.
The main reason
 is that the gauge symmetries do not generate the Lie group action.
This property of $W_N$-gravity is well known \cite{P,LP}.
We consider here in detail the $W_3$ case.
The infinitesimal symmetries are carried out by the second order differential
operators on $\Si_{g,n}$ without constant terms. First we consider
$\SL$-opers \cite{Tel,BD}, which generate the configuration space of $W_3$-gravity.
The action of the second order differential
operators on $\SL$-opers define a Lie algebroid ${\cal A}$ over $\SL$-opers.
The algebroid ${\cal A}$ is lifted to the Hamiltonian algebroid  ${\cal A}^H$ over
the phase space of $W_3$-gravity. The symplectic quotient of the phase space
is the so-called ${\cal W}_3$-geometry of $\Si_{g,n}$. Roughly speaking, this space is
a combination of the moduli of generalized complex structures
 and the spin 2 and 3 fields as the dual variables.
 Note that we deform the operator of complex structure $\bp$ by symmetric
combinations of vector fields $(\ve\p)^2$, in contrast with \cite{BK}, where
the deformations of complex structures are carried out by the polyvector fields.
To define the $W_3$-geometry
we construct the BRST operator for the Hamiltonian algebroid. As it follows from
the general construction, it has the same structure as in the Lie algebra case.
It should be noted that the BRST operator for the $W_3$-algebras was constructed
in \cite{TM}. But here we construct the BRST operator for the different object
- the algebroid symmetries of $W_3$-gravity.  Recently, another BRST description of
$W$-symmetries  was proposed in \cite{BL}.
We explain our formulae and the origin of the algebroid
by the special gauge procedure of the $\SL$ Chern-Simons theory using an
approach developed in \cite{BFK}.

The next example is the Poisson sigma-model \cite{I,SS}.
Cattaneo and Felder \cite{CF1} used this model
for field theoretical explanations of the Kontsevich deformation quantization
formula \cite{K} by the Feynman diagrams technique.
The phase space of the Poisson sigma-model has the same type of Hamiltonian algebroid
 symmetries as in the previous example.
It is worthwhile to note that the Hamiltonian algebroids differ from the
symplectic algebroids that applied in \cite{CF2} to describe the symmetries
of  the Poisson sigma-model in the Lagrangian form. While the Hamiltonian algebroids
describe canonical transformations of symplectic manifolds, the symplectic
algebroids carry symplectic structures by themselves.
We construct
the BRST operator in the Hamiltonian picture. It has the third degree in ghosts
as it should be. This result by no means new. The symmetries and the corresponding BRST
operator were investigated in \cite{SS}.

{\bf Acknowledgments.}\\
{\sl This work was originated from discussions with S. Barannikov
concerning generalized deformations of complex structures during
the visit of the second author the IHES (Bur-sur-Yvette) in 1999.
We benefited also from valuable discussions with  A. Gerasimov, S. Liakhovich and
M. Grigoriev. The work of A.L. is supported in part by grants RFFI-98-01-00344
and 96-15-96455 for support of scientific schools.
The work of M.O. is supported in part by grants RFFI-00-02-16530,
 INTAS 99-01782  and 96-15-96455 for support of scientific schools.
}

\section { Hamiltonian  algebroids and  groupoids}
\setcounter{equation}{0}

We consider in this section   Hamiltonian  algebroids and groupoids.
They are generalizations of the Lie algebras of vector fields.

{\bf 1. Lie algebroids and groupoids}.
We start from a brief description of Lie algebroids and Lie groupoids.
Details of this theory can be find in \cite{KM,Ma,We}.
\begin{defi}
A {\em Lie algebroid} over a differential manifold $V$ is a vector bundle
${\cal A}\rar V$ with
a Lie algebra structure on the space of section $\G({\cal A})$
 defined by the brackets $\lb\ve_1,\ve_2\rb,\\
\ve_1,\ve_2\in\G({\cal A})$ and
a bundle map ({\em the anchor}) $\de :{\cal A}\to TV$, satisfying the following
conditions:\\
(i) For any $\ve_1,\ve_2\in\G({\cal A})$
\beq{5.1}
[\de_{\ve_1},\de_{\ve_2}]=\de_{\lb\ve_1\ve_2\rb},
\eq
(ii) For any $\ve_1,\ve_2\in\G({\cal A})$ and $f\in C^\infty(V)$
\beq{5.2}
\lb\ve_1,f\ve_2\rb=f\lb\ve_1,\ve_2\rb + (\de_{\ve_1} f)\ve_2.
\eq
\end{defi}
In other words, the anchor  defines a representation in the Lie algebra of vector fields
on $V$. The second condition is the Leibniz rule with respect to the multiplication
the sections by smooth functions.

Let $\{e^j(x)\}$ be a basis of local sections $\G ({\cal A})$. Then  the brackets
are defined by the structure functions $f^{jk}_i(x)$ of the algebroid
\beq{5.1b}
\lb e^j,e^k\rb=f^{jk}_i(x)e^i,~~x\in V.
\eq
Using the Jacobi identity for the anchor action, we find
\beq{5.3}
C^n_{j,k,m}\de_{e_n}=0,
\eq
where
\beq{5.4}
C^n_{jkm}=(f^{jk}_i(x)f_n^{im}(x)+\de_{e^m}f^{jk}_n(x)+{\rm c.p.}(j,k,m))
\footnote{The sums over repeated indices are understood throughout the paper.}
\eq
Thus, (\ref{5.3}) implies {\em the anomalous Jacobi identity} (AJI)
\beq{5.5}
f^{jk}_i(x)f_n^{im}(x)+\de_{e^m}f^{jk}_n(x)+{\rm c.p.}(j,k,m)=0
\eq

There exists a global object - the {\em  Lie groupoid} \cite{Ba,KM,Ma,We}.
\begin{defi}
A Lie groupoid $G$ over a manifold $V$
is a pair of differential manifolds $(G,V)$,  two
differential mappings $l,r~: ~G\to V$ and a partially defined binary operation
(a product)
$(g,h)\mapsto g\cdot h $ satisfying the following conditions:\\
(i) It is defined only when $l(g)=r(h)$. \\
(ii) It is associative: $(g\cdot h)\cdot k=g\cdot (h\cdot k)$
whenever the products are defined.\\
(iii) For any $g\in G$ there exist the left and right identity elements $l_g$ and
$r_g$ such that $l_g \cdot g=g\cdot r_g=g$.\\
(iv) Each $g$ has an inverse $g^{-1}$ such that $g\cdot g^{-1}=l_g$ and
$g^{-1}\cdot g=r_g$.\\
\end{defi}
We denote an element of $g\in G$ by the triple $<x|g|y>$, where $x=l(g),~y=r(g)$.
Then  the product $g\cdot h $ is
$$
g\cdot h\rar <x|g\cdot h|z>=<x|g|y><y|h|z>.
$$
An orbit of the groupoid in the base $V$ is defined as an equivalence
$x\sim y$ if $x=l(g),~y=r(g)$. There is the isotropy subgroup $G_x$ for $x\in V$.
$$
G_x=\{g\in G~|~l(g)=x=r(g)\}\sim\{<x|g|x>\}.
$$

The Lie algebroid is a local version of the Lie groupoid.
It is obtained in the following way.
(The details can be found in \cite{Ba}).
Let $f(x|g)=x'$ for $<x|g|x'>$. In terms of $f$ the multiplication
$g\cdot g'$ is defined
by the function $\vf(g,g';x)$ corresponding to the triple $<x|g\cdot g'|x'>$
$$
f'(f(x|g)|g')=f'(x|\vf(g,g';x)) ,~~x'=f(x|\vf(g,g';x)).
$$
Then the anchor takes the form
$$
\de_{e^k}=\frac{\p f^a(x|g)}{\p g_k}|_{g=r_g}\p_{x_a}.
$$
The structure functions are read off from $\vf$:
$$
f^{jk}_i(x)=
(\frac{\p^2}{\p g^j\p h^k}-\frac{\p^2}{\p g^k\p h^j})\vf_i(g,h;x)|_{g=r_g,h=r_h}
$$
It can be proved that (\ref{5.1}) and (\ref{5.5}) provide the reconstruction of the
Lie groupoid from the Lie algebroid at least locally.
\bigskip

{\bf 2.Lie algebroid representations and Lie algebroid cohomology.}
The definition of the algebroids representation is rather evident:
\begin{defi}
A vector bundle representation (VBR) $(\rho, {\cal M})$ of the Lie algebroid
$\cal A$ over the manifold $V$
is a vector bundle $\cal M$ over $V$ and a bundle map
 $\rho$ from $ {\cal A}$ to the bundle of differential operators on $\cal M$ of the order
 less or equal to $1$ $\mbox{\it Diff}^{\le 1}({\cal M}, {\cal M})$,
compatible with the anchor map and commutator such that:\\
(i) the symbol of $\rho(\ve)$ is a scalar equal to the anchor of $\ve$:
$$
{\rm Symb}(\rho(\ve))=\de_\ve{\rm Id}_{\cal M}
$$
(ii) for any $\ve_1,\ve_2\in\G({\cal A})$
\beq{rep}
[\rho({\ve_1}),\rho({\ve_2})]=\rho({\lb\ve_1\ve_2\rb}),
\eq
 where  the l.h.s. denotes
the commutator of differential operators.
\end{defi}

For example, the trivial bundle is a VBR representation (the map $\rho$
is the anchor map $\delta$),

Consider a small disk $U_{\alpha} \subset V$ with local coordinates
$x=(x_1,\ldots,x_a,\ldots)$. Then the anchor can be written as
\beq{5.15i}
\de_{e^j}=b^j_a(x)\frac{\p}{\p x_a}=<b^j|\frac{\de}{\de  x}>.
\footnote{The brackets $<|>$ mean summations over all indices, taking a traces,
integrations, etc.}
\eq
Let $w$ be a section of the tangent bundle $TV$. Then the VBR on $TV$ takes
the form
\beq{5.15j}
{\rho}_{e^j}w=<b|\frac{\de}{\de x}w>-<\frac{\de}{\de x}w|b^j(x)>.
\eq
Similarly, the VBR the action of $\rho$ on a section $p\in T^*V$ is
\beq{5.16i}
\rho_{e^j}p=\frac{\de}{\de x}<p|b^j(x)>.
\eq
We drop a more general definition of the sheaf representation.

We shall define  cohomology groups of  algebroids.
First, we consider the case of contractible base $V$.
Let ${\cal A}^*$ be al bundle over $V$ dual to ${\cal A}$.
 Consider the bundle of graded commutative algebras
$\wedge^\bullet{\cal A}^*$.
The space $\G(V, \wedge^\bullet {\cal A}^*)$
is generated by the sections $\eta_k$:
$<\eta_j,e^k>=\de_j^k$. It is a graded algebra
$$
\G(V, \wedge^\bullet  {\cal A}^*)=\oplus {\cal A}^*_n,~~
{\cal A}^*_n=\{c_n(x)=\f1{n!}c_{j_1,\ldots,j_n}(x)\eta_{j_1}\ldots\eta_{j_n},~x\in V\}.
$$

 Define the Cartan-Eilenberg operator ``dual'' to the  brackets $\lb ,\rb$
$$
sc_n(x;e^{1},\ldots,e^{n},e^{n+1})=
(-1)^{i-1}\de_{e^i}c_n(x;e^{1},\ldots,\hat{e}^{i},e^{n})-
$$
\beq{5.6}
-\sum_{j<i}
(-1)^{i+j}c_n(x;\lb e^{i},e^j\rb,\ldots,\hat{e}^{j},\ldots,\hat{e}^{i},\ldots,e^{n}).
\eq
It follows from (\ref{5.1}) and AJI (\ref{5.5}) that $s^2=0$.
Thus,
$s$ determines a complex of bundles ${\cal A}^*\to \wedge^2{\cal A}^*\to \cdots$.

The cohomology group of this complex are called {\em the cohomology group of
algebroid with trivial coefficients}.
This complex is  a part of the BRST complex
derived below and $\eta$ will play the role of {\em the ghosts}.
 The action of the coboundary operator $s$ takes the following form
 on the lower cochains:
\beq{5.7}
sc(x;\ve)=\de_\ve c(x),
\eq
\beq{5.8}
sc(x;\ve_1,\ve_2)=\de_{\ve_1}c(x;\ve_2)-\de_{\ve_2}c(x;\ve_1)-
c(x;\lb \ve_1,\ve_2\rb)
\eq
\beq{5.9}
sc(x;\ve_1,\ve_2,\ve_3)=\de_{\ve_1}c(x;\ve_2,\ve_3)-
\de_{\ve_2}c(x;\ve_1,\ve_3)
\eq
$$
+\de_{\ve_3}c(x;\ve_1,\ve_2)
-c(x;\lb \ve_1,\ve_2\rb,\ve_3)
+c(x;\lb \ve_1,\ve_3\rb,\ve_2)
-c(x;\lb \ve_2,\ve_3\rb,\ve_1).
$$

It follows from (\ref{5.7}) that $H^0({\cal A},V)$  is isomorphic to the
invariants in the space $C^\infty (V)$. The next cohomology group
$H^1({\cal A},V)$ is responsible for the shift of the anchor action:
\beq{5.11}
\hat{\de}_\ve f(x)= \de_\ve f(x)+c(x;\ve), ~~sc(x;\ve)=0.
\eq
If $c(x;\ve)$ is a cocycle (see (\ref{5.8})), then this action is
 consistent with the defining anchor property (\ref{5.1}). The action of
$\hat{\de}_\ve $ for exact cocycles just gives the shift \\
$\hat{\de}_\ve f(x) =\de_\ve(f(x)+c(x))$.

Instead of $f(x)$
consider $\Psi(x)=\exp f(x)$. The action (\ref{5.11}) on $\Psi$
takes the form
\beq{5.11a}
\hat{\de}_{\ve}\Psi=\de_\ve \Psi(x)+c(x;\ve)\Psi(x).
\eq
This formula defines a ``new'' structure of VBR on the {\em trivial } line
bundle.

 Let $\ti{V}=V/G$ be the
set of orbits of the groupoid $G$ on its base $V$.
The condition
\beq{5.11b}
\hat{\de}_{\ve}\Psi=0
\eq
defines a linear bundle ${\cal L}(\ti{V})$ over $\ti{V}$. In concrete
examples it can be
identified with a determinant bundle over $\ti{V}$.

Two-cocycles $c(x;\ve_1,\ve_2)$ allow to construct
the central extensions of brackets on $\G({\cal A})$
\beq{5.9a}
\lb(\ve_1,k_{1}),(\ve_2,k_{2})\rb_{c.e.}=
(\lb\ve_1,\ve_2\rb,c (x;\ve_1,\ve_2)).
\eq
The cocycle condition (\ref{5.9}) means that the new brackets
$\lb~,~\rb_{c.e.}$ satisfies AJI (\ref{5.5}).
The exact cocycles leads to the splitted extensions.

If $V$ is not contractible the definition of cohomology group is more
complicated. We sketch the \^{C}ech version of it.
Choose an acyclic covering $U_\alpha$. Consider the
corresponding to this covering the \^{C}ech
complex with coefficients in $\bigwedge{}^{\bullet}({\cal A}^{*})$:
$$\bigoplus \Gamma (U_\alpha ,\bigwedge{}^{\bullet}({\cal
A}^{*}))
\stackrel{d}{\longrightarrow}
\bigoplus\Gamma (U_{\alpha\beta },\bigwedge{}^{\bullet}({\cal
A}^{*}))
\stackrel{d}{\longrightarrow}\cdots$$
The \^{C}ech differential $d$ commutes with  the Cartan-Eilenberg operator
$s$, and cohomology of algebroid are cohomology of normalization of this
bicomplex~:
$$\bigoplus \Gamma (U_\alpha ,{\cal
A}^{*}_0)
\stackrel{d,s}{\longrightarrow}
\bigoplus\Gamma (U_{\alpha\beta },{\cal
A}^{*}_0)
\oplus
\bigoplus \Gamma (U_\alpha ,{\cal
A}^{*}_1)
\stackrel{
\left(\begin{array}{ccc}
d&s&\\
&-d&s
\end{array}\right)
}{\longrightarrow}$$
$$
\bigoplus\Gamma (U_{\alpha\beta\gamma },{\cal
A}^{*}_0)
\oplus
\bigoplus \Gamma (U_{\alpha\beta} ,{\cal
A}^{*}_1)
\oplus
\bigoplus \Gamma (U_\alpha ,{\cal
A}^{*}_2) \cdots.
$$
So, the cochains are bigraded
 $c^{i,j}\in\bigoplus_{\alpha_{1}\alpha_{2}\cdots \alpha_j}
\Gamma (U_{\alpha_{1}\alpha_{2}\cdots \alpha_j} ,{\cal
A}^{*}_i)$ and the differential maps $c^{i,j}$ to $(-1)^jd\,c^{i,j}+sc^{i,j}$,
$(-1)^jd\,c^{i,j}$ has type $(i,j+1)$ and  $sc^{i,j}$ has type $(i+1,j)$.

Again, the group $H^0({\cal A},V)$  is isomorphic to the
invariants in the whole space $C^\infty (V)$.

Consider the next groups $H^{(1,0)}({\cal A},V)$  and $H^{(0,1)}({\cal A},V)$.
We have  two components\\ $(c_\alpha(x,\ve),c_{\alpha\beta}(x))$.
They are characterized by the following conditions (see (\ref{5.8}))
 $$
c_\alpha(x;\lb \ve_1,\ve_2\rb)=
\de_{\ve_1}c_\alpha(x;\ve_2)-\de_{\ve_2}c_\alpha(x;\ve_1),
$$
\beq{ch1}
  \de_{\ve}c_{\alpha\beta}(x)=-c_\alpha(x;\ve)+c_\beta(x;\ve),
\eq
\beq{ch2}
 c_{\alpha\gamma}(x)=c_{\alpha\beta}(x)+c_{\beta\gamma}(x).
\eq
While the first component  $c_\alpha(x,\ve)$ comes from the algebroid action
on $U_\alpha$ and define the action of the algebroid on the trivial bundle
(\ref{5.11a}) , the second
component  determines a line bundle ${\cal L}$ on $V$ by the transition functions
$\exp (c_{\alpha\beta})$. The condition (\ref{ch2}) shows that the actions on
the restriction to $U_{\alpha\beta}$
are compatible.

The continuation of the central extension (\ref{5.9a}) from $U_\al$ on $V$ is defined now
by\\
$H^{(j,k)}({\cal A},V), j+k=2$.
There are obstacles to this continuations in $H^{(2,1)}({\cal A},V)$.
We do not dwell on this point.

\bigskip
{\bf 3. Hamiltonian algebroids and groupoids}.
Now consider a vector bundle\\
 ${\cal A}^H\rar {\cal R}$ over a symplectic
manifold ${\cal R}$. Any smooth function $h\in C^\infty({\cal R})$
 gives rise to a vector field
$\de_h$
(the canonical transformations). It is defined by the internal derivation $i_h$ of
the symplectic form $i_h\om=d h$.
 The space $C^\infty({\cal R})$ has
the structure of a Lie algebra with respect to the Poisson brackets
$$
\{h_1,h_2\}=-i_{h_1}dh_2.
$$
The canonical transformations of ${\cal R}$ are determined by the Poisson
brackets with the Hamiltonians
$$
\de_{h} x=\{x,h\}.
$$

Assume that the space of sections $\G({\cal A}^H)$ is equipped by
 the antisymmetric brackets $\lb\ve_1,\ve_2\rb$.
\begin{defi}
${\cal A}^H$ is a {\em Hamiltonian  algebroid} over a symplectic
manifold ${\cal R}$ if there is
a bundle map from ${\cal A}^H$ to the Lie algebra
on $C^\infty({\cal R})$: $\ve\to h_\ve$, (i.e. $f\ve\to fh_\ve$ for
$f\in C^\infty({\cal R})$)
satisfying the following conditions:\\
(i) For any $\ve_1,\ve_2\in\G ({\cal A}^H)$ and $x\in {\cal R}$
\beq{5.12}
\{ h_{\ve_1},h_{\ve_2}\}=h_{\lb\ve_1,\ve_2 \rb}.
\eq\\
(ii) For any $\ve_1,\ve_2\in\G ({\cal A}^H)$ and $f\in C^\infty({\cal R})$
$$
\lb\ve_1,f\ve_2\rb=f\lb\ve_1,\ve_2\rb+\{h_{\ve_1},f\}\ve_2.
$$
\end {defi}
The both conditions are similar to the defining properties of the
Lie algebroids (\ref{5.1}),(\ref{5.2}).
\begin{rem}
In contrast with the Lie algebroids with the bundle map
$$
f\ve\to f\de_\ve,~f\in C^\infty(V),~\ve\in\G({\cal A}^H)
$$
 for the Hamiltonian algebroids
one has the map to the first order differential operators with respect to $f$
$$
f\ve\to f\de_{h_\ve}+h_\ve\de_f.
$$
\end{rem}

Let
$$
C_n^{j,k,m}=f^{jk}_i(x)f_n^{im}(x)+\{h_{e^m},f^{jk}_n(x)\}+{\rm c.p.}(j,k,m).
$$
Then from the Jacobi identity for the Poisson brackets one obtains
\beq{5.13a}
C_n^{j,k,m}h_{\ve^n}=0.
\eq
This identity is similar to (\ref{5.3}) for Lie algebroids.
But now one can add to $C_n^{j,k,m}$  the term proportional to
$E^{j,k,m}_{[ln]}h_{\ve^l}$ without the breaking (\ref{5.13a})
(here $[,]$ means the antisymmetrization).
Thus the Jacobi identity for the Poisson algebra of Hamiltonians yields
\beq{5.14}
f^{jk}_i(x)f_n^{im}(x)+\{h_{e^m},f^{jk}_n(x)\}+
E^{j,k,m}_{[ln]}h_{\ve^l}+c.p.(j,k,m)=0.
\eq
This structure arises in the Hamiltonian systems with the first class constraints
\cite{Ba} and leads to the open algebra of arbitrary rank (see \cite{HT,BFV}).

The important particular case
\beq{5.15}
f^{jk}_i(x)f_n^{im}(x)+\{H_{\ve^m}f^{jk}_n(x)\}+c.p.(j,k,m)=0
\eq
corresponds to the open algebra of rank one similar to the
Lie algebroid (\ref{5.5}). We will call (\ref{5.15}) a simple anomalous
Jacobi identity (SAJI) preserving the notion AJI for the general form (\ref{5.14}).
In this case the Hamiltonian algebroid can be integrated to the Hamiltonian groupoid.
The later is the Lie groupoid with the canonical action. In other words,
the groupoid action preserves the symplectic form on the base ${\cal R}$.

\bigskip
{\bf 4.Symplectic affine bundles over cotangent bundles}.
We shall define below Hamiltonian algebroids over cotangent bundles which
are a special class of symplectic manifold. There exist a slightly more general
symplectic manifold than a cotangent bundle that we shall include in our
scheme as well. It is  an affinization of the cotangent bundle we are going to
define. Let $M$ be a vector space and ${\cal R}$ is a set with an action of
$M$ on ${\cal R}$
$$
{\cal R}\times M\rar {\cal R}:(x,v)\to x+v\in {\cal R}.
$$
\begin{defi}
The set ${\cal R}$ is an {\sl affinization} over $M$ (a {\em  principle homogeneous
space} over $M$) ${\cal R}/M$ if the action is transitive
and exact.
\end{defi}
In other words for any pair $x_1,x_2\in{\cal R}$ there exists $v\in M$ such that
$x_1+v=x_2$, and $x_1+v\neq x_2$ if $v\neq 0$.

This construction is generalized on bundles. Let $E$ be a bundle over $V$ and
$\G(U,E)$ be the linear space of the sections in a trivialization of $E$ over some
 disk $U$.
\begin{defi}
{\em An affinization} ${\cal R}/E$ of  $E$
 is a bundle over $V$ with the space of
local sections $\G(U,{\cal R})$ defined as the affinization over $\G(U,E)$.
\end{defi}
Two affinizations ${\cal R}_1/E$ and ${\cal R}_2/E$ are equivalent if there
exists a bundle map compatible with the action of the corresponding linear spaces.
It can be proved that non-equivalent affinizations are classified by $H^1(V,\G(E))$.

Let $E=T^*V$.
Consider a linear bundle ${\cal L}$ over $V$. The space of connections
Conn$_V({\cal L})$ can be identified with the space of sections
${\cal R}/T^*V$. In fact, for any
connection $\nabla_x, ~x\in U\subset V$ one can define another connection
$\nabla_x+\xi,~\xi\in \G(T^*V)$. Thus, ${\cal R}/T^*V$ can be classified by
the first Chern class $c_1({\cal L})$. The trivial bundles correspond to
$T^*V$.

The affinization ${\cal R}/T^*V$ is
the symplectic space with the canonical form $<dp\wedge du>$. In contrast with $T^*V$
this form is not exact, since $pdu$ is defined only locally.
In the similar way as for $T^*V$,
the space of square integrable sections $L^2(\G({\cal L}))$ plays the role of
the Hilbert space in the prequantization of the affinization ${\cal R}/T^*V$.
For $f\in{\cal R}$ define the hamiltonian vector field $\al_f$ and the
covariant derivative $\nabla(f)_x=i_{\al_f}\nabla_x $   along $\al_f$.
Then the prequantization of ${\cal R}/T^*V$ is determined by the operators
$$
\rho(f)=\f1{i}\nabla(f)_x+f
$$
acting on the space $L^2(\G({\cal L}))$. In particular,
$\rho(p)=\f1{i}\frac{\de}{\de x}$,
$\rho(x)=x$.

The basic example, though for infinitesimal spaces, is the affinizations over
the antiHiggs bundles
\footnote{We use the antiHiggs bundles instead of the standard Higgs bundles for
reasons, that will be clear in Section 4.}.
The antiHiggs bundle ${\cal H}_N(\Si)$ is a cotangent bundle to the space
of connections $\nabla^{(1,0)}=\p+A$ in a vector bundle of rank $N$
over a Riemann curve
$\Si$. The cotangent vector (the antiHiggs field) is $\sln$ valued $(0,1)$-form
$\bar{\Phi}$. The symplectic form on ${\cal H}_N(\Si)$ is
$-\int_\Si\tr(d\bar{\Phi}\wedge dA)$.
The affinizations ${\cal R}^\ka/{\cal H}_N(\Si)$
are the space of connections $(\ka\bp+\bA,\p+A)$ with
symplectic form $\int_\Si\tr(dA\wedge d\bA)$, where $\ka$ parameterizes the
affinizations. The elements of the space Conn$_{(\p+A)}({\cal L})$
giving rise to   ${\cal R}^\ka_{{\rm SL}(N)}/{\cal H}_N(\Si)$ are
\beq{na}
\nabla \Psi=\frac{\de \Psi}{\de A}+\ka \bA\Psi.
\eq

\bigskip
{\bf 5. Hamiltonian algebroids related to Lie algebroids.}
Now we are ready to introduce an important subclass of Hamiltonian algebroids.
They are extensions of the Lie algebroids and share with them
SAJI (\ref {5.15}) without additional terms as in (\ref{5.14}).
Our two basic examples belong to this subclass.

\begin{lem}
The anchor action (\ref{5.15i}) of the Lie algebroid ${\cal A}$
can be lifted to the Hamiltonian action on ${\cal R}/T^*V$
in a such way that it defines the Hamiltonian algebroid ${\cal A}^H$ over ${\cal R}$.
The equivalence classes of these lifts
are isomorphic to $H^1({\cal A},V)$.
\end{lem}
{\sl Proof}.
Consider a small disk $U_\al\subset V$. The anchor (\ref{5.11}) has the form
\beq{5.15a}
\hat{\de}_{e^j}=<b^j|\frac{\de}{\de  x}>+c(x;e^j).
\eq
Next, continue the action on ${\cal R}/T^*U_\al$.
We represent the affinization
as the space \\
Conn${\cal L}(V)=\{\nabla^p\}$.
Since ${\cal L}$ on $U_{\alpha}$ is trivialized we can identify the connections
with one-forms $p$.
Let $w\in TV$ and
$$
\nabla^p_w\Psi:=i_w\nabla^p\Psi=<w|\frac{\de\Psi}{\de x}>+<w|p>\Psi,~~ x\in U_\alpha,
~~p\in T^*V
$$
be the covariant derivative along $w$.
 To lift the action we use the
Leibniz
rule for the anchor action on the covariant derivatives:
$$
\hat\de_{e^j}(\nabla^p)_\al\Psi=\hat\de_{e^j}(\nabla^p_\al\Psi)-
\nabla^p_\al\hat\de_{e^j}\Psi-\nabla^p_{\hat\de_{e^j}\al}\Psi.
$$
It follows from (\ref{5.15j}),(\ref{5.16i}) and (\ref{5.15a}) that
\beq{5.16a}
\hat{\de}_{e^j}p=-\frac{\de}{\de x}<p|b^j(x)>-\frac{\de}{\de x}c(x;e^j).
\eq
Note that the second term is responsible for the pass from $T^*V$ to
the affinization ${\cal R}$, otherwise $p$ is transformed as a cotangent vector
(see (\ref{5.15i}).

The vector fields $<b^j|\frac{\de}{\de  x}>$ and (\ref{5.16a}) are hamiltonian
 with respect
to the canonical symplectic form $<dp|dx>$ on ${\cal R}$.
The corresponding  Hamiltonians have the linear dependence on "momenta":
\beq{5.17}
h^j=<p|b^j(x)>+c^j(x).
\eq
Note that if $h^j$ satisfies the Hamiltonian algebroid
property (\ref{5.12}), then $sc^j(x)=0$ (\ref{5.8}).

We have constructed the Hamiltonians locally and want to prove that this definition
is compatible with gluing
$U_\alpha$ and $U_\beta$. Note, that when we glue  ${\cal R}|_{U_\alpha}$
 and ${\cal R}|_{U_\beta}$
we shift fibers by $\frac{\de c_{\alpha\beta}}{\de x}$~:
$p_\alpha=p_{\beta}+ \frac{\de c_{\alpha\beta}}{\de x}$. Indeed,
 we glue the bundle ${\cal L}(V)$
restricted on $U_{\alpha\beta}$ by multiplication on  $\exp  c_{\alpha\beta}(x)$.
The connections are
transformed
by adding the logarithmic derivative of the transition functions.
On the other hand, $ \de_{\ve}c_{\alpha\beta}(x)=-c_\alpha(x;\ve)+c_\beta(x;\ve)$
(see (\ref{ch1})). So
$$
h_\alpha^j=<p_\alpha|b^j(x)>+c_\alpha^j(x)=<p_{\beta}+
 \frac{\de c_{\alpha\beta}}{\de x}|b^j(x)>
 -\de_{\ve}c_{\alpha\beta}(x)+c_\beta(x;\ve)$$
$$=<p_\beta|b^j(x)>+c_\beta^j(x)=h_\beta^j,$$
and the Hamiltonians become defined globally.

The exact cocycle $(c_\alpha^j(x)=\de_{e^j}f_\alpha(x),
c_{\alpha\beta}(x)=f_\beta(x)-f_\alpha(x))$
just shifts the momenta
 on the derivative of $f_\alpha(x)$
$$
h^j=<p_\alpha+\frac{\de f_\alpha(x)}{\de x}|b^j(x)>.
$$

We rewrite the canonical transformations in the form
$$
\hat{\de}_{e^j}\Phi(p,x)=\de_{e^j}\Phi(p,x)+
<f^j|\frac{\de\Phi(p,x)}{\de p}>,
$$
$$
f^j=-<p|\frac{\de \be^j(x)}{\de x}>-\frac{\de c^j(x)}{\de x}.
$$
Thus, all nonequivalent
lifts of the anchor $\de$ from $V$ to ${\cal R}/T^*V$ are in one-to-one correspondence
with $H^1({\cal A},V)$.
Thereby, we have constructed the Hamiltonian
algebroid ${\cal A}^H$ over the principle homogeneous  space ${\cal R}$.
It has the same fibers and
the same structure functions $f^{jk}_i(x)$ as the underlying Lie algebroid ${\cal A}$
along with the bundle map $e^j\to h^j$ (\ref{5.17}).
$\Box$

Now investigate AJI (\ref{5.14}) in this particular case.
\begin{lem}
The Hamiltonian algebroids  ${\cal A}^H$ have SAJI (\ref{5.15}).
\end{lem}
{\sl Proof}.
First note, that the Lie algebroids we started from have
SAJI (\ref{5.5}). The Hamiltonian algebroids
${\cal A}^H$ have the same structure functions
$f^{jk}_i(x)$ depending on coordinates on $V$ only. Consider
the general AJI (\ref{5.14}). It follows from (\ref{5.17})
that
$$
\{h^j,f^{nk}_i(x)\}=<b^j|\frac{\de f^{nk}_i(x)}{\de  x}>=
\de_{e^j}f^{nk}_i(x).
$$
The sum of the first two terms in (\ref{5.14}) coincides with
the SAJI (\ref{5.15}) in the underlying Lie algebroid, and therefore vanishes.
$\Box$

\bigskip
{\bf 6. Reduced phase space and its BRST description.}
In what follows we shall consider Hamiltonian algebroids related to
Lie algebroids. Let $e^j$ be a basis of sections in $\G({\cal A}^H)$.
Then the Hamiltonians (\ref{5.17}) can be represented in the form
$h^j=<e^j|F(x)>$, where $F(x)\in\G({\cal A}^{H*})$ defines the {\em moment map}
$$
m: {\cal R}\rar\G({\cal A}^{H*}),~~m(x)=F(x).
$$
The coadjoint action ${\rm ad}^*_{\ve}$ in $\G({\cal A}^{H*})$ is defined
in the usual way
$$
<\lb\ve, e^j\rb|F(x)>=<e^j|{\rm ad}^*_{\ve}F(x)>.
$$
One can fix a moment  $F(x)=m_0$ in $\G({\cal A}^{H*})$.
 The reduced phase space is defined as the quotient
$$
{\cal R}^{red}=\{x\in {\cal R}|(F(x)=m_0)/G_0\},
$$
where $G_0$ is  generated by the  transformations
${\rm ad}^*_\ve$ such that ${\rm ad}^*_\ve m_0=0$. In other words, ${\cal R}^{red}$
is the set of orbits of $G_0$ on the constraint surface $F(x)=m_0$.
The symplectic form $\om$ being restricted on $R^{red}$ is non-degenerate.

The BRST approach allows to go around the reduction procedure by introducing
additional fields (the ghosts).
We shall construct the BRST complex for ${\cal A}^{H}$ in a similar way as the
Cartan-Eilenberg complex for the Lie algebroid ${\cal A}$. In contrast with the
Cartan-Eilenberg complex  the BRST complex has a Poisson structure which
allows to define the nilpotent operator (the BRST operator) in a simple way.

 Consider the dual bundle ${\cal A}^{H*}$.
Its  sections $\eta \in\G({\cal A}^{H*})$ are the odd fields called
 {\em the ghosts}.
Let $h_{e^j}=<\eta_j|F(x)>$, where $\{\eta_j\}$ is a basis in
$\G({\cal A}^{H*})$ and $F(x)=0$ are the moment constraints, generating the
canonical algebroid action on ${\cal R}$.
 Introduce another type of odd variables ({\em the ghost momenta})
${\cal P}^j,~j=1,2,\ldots$ dual to the ghosts $\eta_k,~k=1,2,\ldots$
${\cal P}\in \G({\cal A}^{H})$.
 We attribute the ghost
 number one to the ghost fields gh$(\eta)=1$, minus one to the ghost momenta
gh$({\cal P})=-1$ and gh$(x)=0$ for $x\in {\cal R}$.
Introducing
the Poisson brackets in addition to the non-degenerate Poisson structure on ${\cal R}$
\beq{5.19}
\{\eta_j,{\cal P}^k\}=\de_j^k,~~\{\eta^j,x\}=\{{\cal P}_k,x\}=0.
\eq

Thus all fields are incorporated in the graded Poisson superalgebra
$$
{\cal BFV}=\left(
\G(\wedge^\bullet  ({\cal A}^{H*}\oplus{\cal A}^{H})
\right)\otimes C^\infty({\cal R}).
=\G(\wedge^\bullet {\cal A}^{H*})
\otimes\G(\wedge^\bullet {\cal A}^{H})\otimes C^\infty({\cal R}).
$$
({\em the Batalin-Fradkin-Volkovitsky (BFV) algebra}).

There exists a nilpotent operator on $Q,~Q^2=0,~gh(Q)=1$
({\em the BRST operator}) transforming ${\cal BFV}$ into the BRST complex.
The cohomology of ${\cal BFV}$ complex give rise to the structure of
the classical reduced phase space ${\cal R}^{red}$. In some cases
$H^j(Q)=0,~j>0$
and $H^0(Q)=${classical observables}.

Represent the action of $Q$ as the Poisson brackets:
$$
Q\psi=\{\psi,\Om\},~~\psi,\Om\in {\cal BFV}.
$$
Due to the Jacobi identity for the Poisson brackets the nilpotency of $Q$
is equivalent to
$$
\{\Om,\Om\}=0.
$$
Since $\Om$ is odd, the brackets are symmetric.
 For generic Hamiltonian algebroid $\Om$ has the form \cite{HT}
$$
\Om=h_\eta+\oh<\lb\eta,\eta'\rb|{\cal P}>+...,~(h_\eta=<\eta|F>),
$$
where the terms of order two and more in ${\cal P}$ omit.
The order of ${\cal P}$
in $\Om$ is called {\em the rank} of the BRST operator $Q$.
If ${\cal A}$ is a Lie algebra defined along with its canonical
action on ${\cal R}$ then $Q$ has the rank one or less. In this case
the BRST operator $Q$ is the extension of the Cartan-Eilenberg operator
giving rise to the cohomology of ${\cal A}$ with coefficients in
$C^\infty({\cal R})$ and the first two terms in the previous expression provide
the nilpotency of $Q$. It turns out that  $\Om$ has the same structure
for Hamiltonian algebroids ${\cal A}^H$, though the Jacobi identity has additional
terms in compare with the Lie algebras.
\begin{theor}
The BRST operator $Q$ for the Hamiltonian algebroid ${\cal A}^{H}$
has the rank one:
\beq{5.20}
\Om=<\eta|F>+\oh<\lb\eta,\eta'\rb|{\cal P}>
\eq
\end{theor}

{\sl Proof of theorem}.\\
We use the Poisson brackets coming from the symplectic structure on ${\cal R}$
and (\ref{5.19}). Straightforward  calculations show that
$$
\{\Om,\Om\}=
\{h_{\eta_1},h_{\eta_2}\}+\oh<\lb\eta_2,\eta_2'\rb|F>-
$$
$$
-\oh<\lb\eta_1,\eta_1'\rb|F>
+\oh\{h_{\eta_1},<\lb\eta_2\eta_2'\rb|{\cal P}_2>\}-
$$
$$
\oh\{h_{\eta_2},<\lb\eta_1,\eta_1'\rb|{\cal P}_1>\}+
\f1{4}\{<\lb\eta_1,\eta_1'\rb|{\cal P}_1>,<\lb\eta_2,\eta_2'\rb|{\cal P}_2> \}.
$$
The sum of the first three terms vanishes due to (\ref{5.12}).
The sum of the rest terms is the left hand side of SAJI (\ref{5.15}).
The additional dangerous term may come from the Poisson brackets of
the structure functions
$\{\lb\eta_1,\eta_1'\rb,\lb\eta_2,\eta_2'\rb\}$. In fact, these
brackets vanish because the structure functions
do not depend
on the ghost momenta. Thus the SAJI leads to the desired identity $\{\Om,\Om\}=0$.
$\Box$

\section{Two examples of Hamiltonian Lie algebra symmetries}
\setcounter{equation}{0}

In this section we consider two examples, where the Hamiltonian
algebroids are just the Lie algebras of hamiltonian
vector fields and therefore the symmetries are the standard Lie symmetries.
Nevertheless, they are in much the same
as in the  algebroid cases.  Let $\Si_{g,n}$ be a Riemann curve of genus $g$
with $n$ marked points. The first examples is the moduli
space of flat bundles over $\Si_{g,n}$. It will be clear later, that it is
an universal system containing hidden algebroid symmetries. The second example is
the projective structures (${\cal W}_2$-structures) on $\Si_{g,n}$.
Their generalization is the ${\cal W}_3$-structures, where the symmetries are
defined by a genuine Hamiltonian  algebroid, will be considered in next Section.

{\bf 1.  Flat bundles  with the Fuchsian singularities. }
Consider a ${\rm SL}(N,{\bf C})$ holomorphic bundle  $E$ over a Riemann
curve $\Si_{g,n}$ of genus $g$ with $n$ marked points.
Locally on a disk the connections  $\nabla: E\to E\otimes\Om^{(1,0)}(\Si_{g,n})$,
$\bar{\nabla}: E\to E\otimes\Om^{(0,1)}(\Si_{g,n})$ take the form
\beq{8.0}
\nabla=\p +A, ~~\bar{\nabla}=\bp.
\eq
We assume that $A$ has first order poles at the marked points
\beq{8.0a}
A|_{z\to x_a}=\frac{A_a}{z-x_a}.
\eq
In addition, we consider a collection of $n$  elements from  the
Lie coalgebra
$$
\bfp=(p_1,\ldots,p_a,\ldots,p_n), ~~p_a\in{\rm sl}^*(N,{\bf C}).
$$
Let ${\cal G}_{{\rm SL}(N)}$ be the algebra of the gauge transformations.
It is a $C^\infty$ map
$\Si_{g,n}\to \sln$, or the space of sections of
the bundle $\Om^0(\Si_{g,n},{\rm End}E)$.  Assume that near the marked points
\beq{8.0b}
\ve|_{z\to x_a}=r_a+O(z-x_a), ~~r_a\neq 0.
\eq

The gauge  transformations act as
\beq{8.1}
\de_\ve A=\p\ve+[A,\ve], ~~\de_\ve p_a=[p_a,r_a],~~\ve\in {\cal G}_{{\rm SL}(N)}.
\eq

We have a trivial principle bundle
${\cal A}_{{\rm SL}(N)} ={\cal G}_{{\rm SL}(N)}\times V_{{\rm SL}(N)}$
over \\
$V_{{\rm SL}(N)}=\{\nabla=\p+A,\bfp\}$   with sections $\ve$ endowed
with the standard matrix commutator
and the anchor map (\ref{8.1}).

The cohomology
$H^i({\cal A}_{{\rm SL}(N)})=H^i({\cal G_{{\rm SL}(N)}},V_{{\rm SL}(N)})$
are the standard cohomology
of the gauge algebra ${\cal G}_{{\rm SL}(N)}$ with  cochains taking values
in functionals on $V_{{\rm SL}(N)}$.  There is a nontrivial one-cocycle
\beq{8.2}
c(A;\ve)=\int_{\Si_{g,n}}\tr
\left(
\ve(\bp A-2\pi i\sum_{a=1}^n\de(x_a)p_a
\right)=
<\ve|\bp A>-2\pi i\sum_{a=1}^n\tr(r_a\cdot p_a)
\eq
representing an element of $H^1({\cal A}_{{\rm SL}(N)})$.
Here $\de(x_a)$ is the functional on $C^\infty(V)$\\
$
f(x_a)=\int_{\Si_{g,n}}f\de(x_a).
$
The contribution of the marked points is equal to
$$
2\pi i\sum_{a=1}^n\tr(r_a(A_a-p_a)).
$$
It follows from (\ref{5.8}) and (\ref{8.1}) that the coboundary operator $s$
annihilates $c(A;\ve)$.
On the other hand, $c(A;\ve)\neq\de_\ve f(A)$ for any $f(A)$.
 This cocycle provides the nontrivial
extension of the anchor action (see (\ref{5.11})
\beq{8.2a}
\hat{\de}_\ve f(A,\bfp)= \de_\ve f(A,\bfp)+
\int_{\Si_{g,n}}\tr(\ve\bp A)-2\pi i\sum_{a=1}^n\tr(r_a\cdot p_a)=
\eq
$$
=\int_{\Si_{g,n}}\tr\left[\frac{\de f}{\de A}(\p\ve+[A\ve])+\ve\bp A\right]
-2\pi i\sum_{a=1}^n\tr(r_a\cdot p_a)=
$$
$$
=\int_{\Si_{g,n}}\tr\ve
\left(\bp A-\p\frac{\de f}{\de A}+[\frac{\de f}{\de A},A]
-2\pi i\sum_{a=1}^n\de(x_a) p_a \right).
$$

Next consider $2g$ contours $\ga_\al,~\al=1,\ldots,2g$
 generating $\pi_1(\Si_{g})$. The contours determine  2-cocycles
\beq{8.3}
c_\al(\ve_1,\ve_2)=\int_{\ga_\al}\tr(\ve_1\p\ve_2).
\eq
Due to the smooth behavior of the gauge algebra at the marked points (\ref{8.0b}),
the contour integrals around them vanish.
The cocycles (\ref{8.3}) allow to construct $2g$ central extensions
$\hat{\cal G}_{{\rm SL}(N)}$ of ${\cal G}_{{\rm SL}(N)}$
$$
\hat{\cal G}_{{\rm SL}(N)}={\cal G}_{{\rm SL}(N)}\oplus_{\al=1}^{2g}{\bf C}\La_\al,
$$
$$
[(\ve_1,\sum_{\al}k_{1,\al}),(\ve_2,\sum_{\al}k_{2,\al})]_{c.e.}=
\left([\ve_1,\ve_2],\sum_{\al}c_\al(\ve_1,\ve_2)\right).
$$

Consider the cotangent bundle $T^*V_{{\rm SL}(N)}$. The conjugate to
$\p +A$ variables are the one-forms
$\bar{\Phi}\in\Om^{(0,1)}(\Si_{g,n},{\rm sl}(N,{\bf C})$ - the antiHiggs field.
In fact, we shall consider the affinization ${\cal R}_N$ over
$T^*V_{{\rm SL}(N)}$ provided
by the cocycle (\ref{8.2}). We have already mentioned that
 the role of momenta plays by  the holomorphic connection $\bp+\bA$ (\ref{na}).
 We put $\ka=1$.
  The dual variables at the marked points are constructed by means of group elements
 $g_a\in{\rm SL}(N,{\bf C})$
$$
\bfg=(g_1,\ldots,g_a,\ldots,g_n), ~~g_a\in{\rm SL}(N,{\bf C}).
$$

The symplectic form on ${\cal R}_{{\rm SL}(N)}$ is
\beq{8.4a}
\om= \int_{\Si_{g,n}}\tr(dA\wedge d\bA)+\sum_{a=1}^n\om_a,
\eq
where
$
\om_a=\tr(d(p_ag_a^{-1})\wedge dg_a),
$
and $dA,~d\bA$ are variations of fields.
In fact, each $\om_a$ is degenerate. It becomes non-degenerate
by the restrictions on the coadjoint orbits
$$
{\cal O}_a=\{p_a=g_a^{-1}p_a^{(0)}g_a~|~p_a^{(0)}=\di(\la_{a,1},\ldots,\la_{a,N}),~
\la_{a,j}\neq\la_{a,k},~g_a\in{\rm SL}(N,{\bf C})\},
$$
where $\om_a$ coincides with the Kirillov-Kostant form
$\om_a=\tr(p^{(0)}_adg_adg_a^{-1})$.

 According to
(\ref{5.17}) the lift of the anchor (\ref{8.1}) to ${\cal R}_{{\rm SL}(N)}$,
defined by the cocycle $c(A;\ve)$ (\ref{8.2}) leads  to the Hamiltonian
$$
h_\ve=\int_{\Si_{g,n}}\tr
\left[
\bA(\p\ve+[A,\ve])+\tr(\ve\bp A)
-2\pi i\sum_{a=1}^n\de(x_a)\ve p_a)
\right]=
$$
$$
<\ve|F(A,\bA)-2\pi i\sum_{a=1}^n\de(x_a)p_a)>,~~
F(A,\bA)=\bp A-\p\bA+[\bA,A].
$$
It generates the canonical vector fields (\ref{8.1}) and
$$
\de_\ve \bA=\bp\ve+[\bA,\ve],~~\de_\ve g_a=g_ar_a,
$$
(see (\ref{5.16a})).
The global version of this transformations is the gauge group
$G_{{\rm SL}(N)}$ acting on the affinization ${\cal R}_{{\rm SL}(N)}$ over
$T^*V_{{\rm SL}(N)}$.
The flatness condition
\beq{8.5a}
m:=F(A,\bA)-2\pi i\sum_{a=1}^n\de(x_a)p_a=0
\eq
is the moment constraint with respect to this action. This equation means
that the residues $A_a$ of $A$ in the marked points (\ref{8.0a}) coincide with
$p_a$.
The flatness is the compatibility condition for the linear
system
\beq{8.5c}
\left\{
\begin{array}{c}
(\p +A)\psi=0,\\
(\bp +\bA)\psi=0,
\end{array}
\right.
\eq
where $\psi\in\Om^0(\Si_{g,n},{\rm Aut} E)$. The second equation
describes the deformation of the holomorphic structure of the bundle $E$
 (\ref{8.0}).

The moduli space ${\cal M}^{flat}_N$ of flat ${\rm SL}(N)$-bundles
is the symplectic quotient\\
${\cal R}_{{\rm SL}(N)}//G_{{\rm SL}(N)}$. It has  dimension
\beq{8.5b}
\dim {\cal M}^{flat}_N=2(N^2-1)(g-1)+N(N-1)n,
\eq
where the last term is the contribution of the coadjoint orbits ${\cal O}_a$
located at the marked points.
Let $G_{{\rm SL}(N)}$ be the gauge group and
$\ti{V}_{{\rm SL}(N)}=V_{{\rm SL}(N)}/G_{{\rm SL}(N)}$ be the set of the
gauge orbits.
Consider a smooth functional $\Psi(A,\bfp)$ on $V_{{\rm SL}(N)}$ such that
$$
\hat{\delta}_\varepsilon\Psi(A,\bfp):=\int_{\Si_{g,n}}\tr\ve\left(
-\p\frac{\de \Psi}{\de A}+[\frac{\de \Psi}{\de A},A]
\right)
 +\left(
\int_{\Si_{g,n}}\tr\ve\left(\bp A-2\pi i\sum_{a=1}^n\de(x_a) p_a\right)
\right)\Psi=0.
$$
These functionals generate the space of sections of the linear bundle
${\cal L}(\ti{V}_{{\rm SL}(N)})$ we
discussed before (\ref{5.11b}).
The linear bundle ${\cal L}$ over $\ti{V}_{{\rm SL}(N)}$  is the determinant bundle
 $\det(\p+A)$ \cite{ADW,H1}.
The prequantization of ${\cal M}^{flat}_N$ is defined in the Hilbert space
of $\G({\cal L}(\ti{V}_{{\rm SL}(N)}))$.

On the other hand ${\cal M}^{flat}_N$ can be described by the
cohomology $H^k(Q)$ of the BRST operator $Q$ which we are going to define.
Let $\eta$ be the  dual to $\ve$ fields (the ghosts)
and ${\cal P}$ are their momenta ${\cal P}\in\Om^{(1,1)}(\Si_{g,n},{\rm End}E)$.
 Consider the algebra
$$
C^\infty({\cal R}_{{\rm SL}(N)})\otimes\wedge^\bullet
 \left({\cal G}_{{\rm SL}(N)})\oplus {\cal G}^*_{{\rm SL}(N)}\right).
$$
Then the BRST operator $Q$ acts on functionals on this algebra as
$$
Q\Psi(A,\bA,\eta,{\cal P})=\{\Om,\Psi(A,\bA,\eta,{\cal P})\},
$$
 where
$$
\Om=<\eta|F(A,\bA)>+\oh<[\eta,\eta']|{\cal P}>=
\int_{\Si_{g,n}}\tr\left(\eta(\bp A-\p\bA+[\bA,A])\right)
+\oh\tr([\eta,\eta']{\cal P}),
$$
where res$A|_{x_a}=p_a$.
\bigskip

{\bf 2. Projective structures on $\Si_{g,n}$}.
Let us fix a complex structure on  $\Si_{g,n}$ by choosing  local coordinates
$(z,\bz)$ and the corresponding operators $\bp$.
Consider the
projective connection $T$  on $\Si_{g,n}$. It corresponds to
the second order differential operator $\p^2-T$ on a disk.
Under the holomorphic diffeomorphisms $T$ is transformed as
$(2,0)$-differential up to addition the Schwarzian derivative.
It means that locally the action of a smooth  vector field
$\ve=\ve(z,\bz)\frac{\p}{\p z}$ on $T$ has the form
\beq{8.7}
\de_{\ve}T(z,\bz)=-\ve\p T-2T\p\ve-\frac{1}{2}\p^3\ve.
\eq
We assume that $T$ has poles at the marked points $x_a,(a=1,\ldots,n)$
  up to the second order:
\beq{8.8a}
T|_{z\rar x_a}\sim\frac{T^a_{-2}}{(z-x_a)^2}+\frac{T^a_{-1}}{(z-x_a)}+\ldots.
\eq
The vector fields generate the Lie algebra ${\cal G}_1$ of the first order differential
operators on $\Si_{g,n}$ with respect to the brackets
$$
[\ve_1,\ve_2]=\ve_1\p\ve_2-\ve_2\p\ve_1,
$$
and the vector fields have the first order holomorphic nulls at the marked points
\beq{8.8b}
\ve|_{z\rar x_a}=r_a(z-x_a)+o(z-x_a).
\eq
We consider the affine space of the projective connection
$V_2=\{\p^2-T\}$ as the base
 of the trivial Lie algebroid ${\cal A}_2$ with the space of sections
 ${\cal G}_1=\{\ve\}$. The anchor is defined by (\ref{8.7}).

Consider the cohomology $H^\bullet ({\cal A}_2)\sim H^\bullet ({\cal G}_1,V_2)$.
Because of (\ref{8.7}) and (\ref{8.8b}) $T^a_{-2}$ in (\ref{8.8a}) represents an element
from $H^0({\cal A}_2)$
\beq{8.8c}
\de_{\ve}T^a_{-2}=0.
\eq

The anchor action (\ref{8.7}) can be extended by the one-cocycle $c(T;\ve)$
representing a nontrivial element of $H^1({\cal A}_2)$
\beq{8.9}
\hat{\de}_{\ve}f(T)=\int_{\Si_{g,n}}\left (
\de_{\ve}T\frac{\de f(T)}{\de T}\right)+c(T;\ve)
,~
c(T;\ve)=\int_{\Si_{g,n}}\ve\bp T,
\eq
The contribution of the marked point in this cocycle is $2\pi ir_aT_{-2}^a$.

As in general case one can consider the quotient space $\ti{V}_2=V_2/G_1$,
where $G_1$ is the group corresponding to the algebra ${\cal G}_1$.
 The space of sections of
the linear bundle ${\cal L}(\ti{V}_2)$ is defined as the space of
functionals $\{\Psi(T)\}$ on $V_2$ that satisfy the following condition
$$
\hat{\delta}_\ve\Psi(T):=\int_{\Si_{g,n}}\ve\left ((\oh\p^3+2T\p+\p T)\frac{\de \Psi}{\de T}
+\bp T\Psi\right)=0.
$$
The linear bundle ${\cal L}(\ti{V}_2)$ is the determinant line bundle
$\det(\p^2-T)$ considered in \cite{Za,Mat}.

There are $2g$ nontrivial two-cocycles defined by the integrals
over non contractible contours $\ga_\al$:
$$
c_\al (\ve_1,\ve_2)=\oint_{\ga_\al}\ve_1\p^3\ve_2.
$$
They give rise to the central extension $\hat{\cal G}_1$ of the Lie algebra
of the first order differential operators on $\Si_g$.

The affinization ${\cal R}_2$ over the cotangent bundle $T^*V_2$ has
the Darboux coordinates
$T$ and $\mu$,
where $\mu\in\Om^{(-1,1)}(\Si_{g,n})$ is the Beltrami differential.
The anchor (\ref{8.7}) is lifted to ${\cal R}_2$ as
\beq{8.11}
\de_\ve\mu=-\ve\p\mu +\mu\p\ve+\bp\ve,
\eq
where the last term occurs due to the cocycle (\ref{8.9}).
We specify the dependence of $\mu$ on the positions of
the marked points in the following  way. Let ${\cal U}'_a$ be a neighborhood
 of the marked point $x_a,~(a=1,\ldots,n)$
such that ${\cal U}'_a\cap{\cal U}'_b=\emptyset$ for $a\neq b$.
Define a smooth function $\chi_a(z,\bz)$
\beq{cf}
\chi_a(z,\bz)=\left\{
\begin{array}{cl}
1,&\mbox{$z\in{\cal U}_a$ },~{\cal U}'_a\supset{\cal U}_a\\
0,&\mbox{$z\in\Si_g\setminus {\cal U}'_a.$}
\end{array}
\right.
\eq
Due to (\ref{8.11}) at the neighborhoods of the marked points $\mu$ is defined
 up to the term\\ $\bp(z-x_a)\chi(z,\bz)$.
Then $\mu$ can be represented as
$$
 \mu=\sum_{a=1}^n[t^{(1)}_{0,a}+t^{(1)}_{1,a}(z-x_a)+\ldots]\mu^0_a,~~
\mu^0_a=\bp \chi_a(z,\bz),~~(t_{0,a}=x_a-x_a^0),
$$
where only $t_{0,a}$ can not be removed by the gauge transformations (\ref{8.11}).
The symplectic form on ${\cal R}_2$ is
$$
\om=\int_{\Si_{g,n}}dT\wedge d\mu.
$$
For rational curves $\Si_{0,n}$ it takes the form
\beq{8.14}
\om=dT^a_{-2}\wedge dt_{1,a}+dT^a_{-1}\wedge dt_{0,a}.
\eq
\begin{rem}
The space ${\cal R}_2$ is the classical phase space
of the $2+1$-gravity on $\Si_{g,n}\times I$ \cite{Ca}. In fact,
$\mu$ is related to the conformal class of metrics on $\Si_{g,n}$
and plays the role of a coordinate, while $T$ is a momentum.
In our construction $\mu$ and $T$ interchange their roles.
\end{rem}

The Hamiltonian of the canonical transformations has the form
\beq{8.12}
h_\ve=\int_{\Si_{g,n}}\mu\de_\ve T+c(T,\ve)=\int_{\Si_{g,n}}\ve F(T,\mu),
\eq
$$
F(T,\mu)=
(\bp+\mu\p+2\p\mu)T-\frac{1}{2}\p^3\mu.
$$

The moment map  $m:{\cal R}_2\rar {\cal G}^*_1$ has the form
\beq{8.13}
m=(\bp+\mu\p+2\p\mu)T-\frac{1}{2}\p^3\mu,
\eq
where ${\cal G}^*_{1}$ is the dual space to the algebra ${\cal G}_{1}$
of vector fields.
${\cal G}^*_{1}$ is the space of distributions  of $(2,1)$-forms on $\Si_{g,n}$.
As it follows from
(\ref{8.8b}) in the neighborhoods of the marked points the elements
$y\in{\cal G}^*_{1}$ take the form
\beq{8.13a}
y\sim b_{1,a}\p\de(x_a)+b_{2,a}\p^2\de(x_a)+\ldots.
\eq
We take
\beq{8.15}
F(T,\mu)=m,~~m=-\sum_{a=1}^nT^a_{-2}\p\de(x_a).
\eq
The algebra  ${\cal G}_1$  preserves $m~:~~{\rm ad}_\ve^*m=m$
for any $\ve$.
Thus, in contrast with the previous example, we have trivial coadjoint orbits at
the marked points. Since $T^a_{-2}$ are fixed the dynamical parameters  are
$t_{0,a},T^a_{-1}$ that contribute in the symplectic structure (\ref{8.14})
The moduli space ${\cal W}_2$ of projective structure on $\Si_{g,n}$ is
the symplectic quotient of ${\cal R}_2$ with respect to the action of $G_1$
$$
{\cal W}_2={\cal R}_2//G_1=\{F(T,\mu)-m=0\}/G_1.
$$
It has dimension $6(g-1)+2n$.

Let $\psi$ be a $(-\oh,0)$ differential. Then (\ref{8.15})
is the compatibility condition for the linear system
\beq{8.16}
\left\{
\begin{array}{l}
(\p^2-T)\psi=0,\\
(\bp+\mu\p -\oh\p\mu) \psi=0.
\end{array}
\right.
\eq
It follows from the second equation that the Beltrami differential $\mu$ provides
 the deformation of complex structure on $\Si_{g,n}$. Note, that we started
from the first equation defining the projective
connection  and $\bp\psi=0$ on $V_2$. The second equation in (\ref{8.16}) emerges
after the passage from $V_2$ to  ${\cal R}_2$ by means of the cocycle (\ref{8.9}).

The tangent space ${\cal T}_2$ to ${\cal W}_2$ is isomorphic
to the cohomology $H^0$ of the BRST complex. It is generated by
the fields $T,\mu\in {\cal R}_2$, the ghosts fields $\eta$ dual to the
vector fields $\ve$ acting via the anchor (\ref{8.7}),(\ref{8.11})
on ${\cal R}_2$ and the ghosts momenta ${\cal P}$.
The BRST operator $Q$ is defined by $\Om$ $(Qf=\{f,\Om\})$
$$
\Om=\int_{\Si_{g,n}}\eta F(T,\mu)+\oh\int_{\Si_{g,n}}[\eta,\eta']{\cal P}.
$$
The first term is just the Hamiltonian (\ref{8.12}), where the vector fields are
replaced by the ghosts.

\section{Hamiltonian algebroid structure in ${\cal W}_3$-gravity}
\setcounter{equation}{0}

Now consider the concrete example of the general construction with nontrivial
algebroid structure.
It is the ${\cal W}_3$-structures on $\Si_{g,n}$ \cite{P,BFK,GLM}
which generalize the projective structures described in previous Section.

{\bf 1. $\SL$-opers.}
Opers are $G$-bundles over Riemann curves with additional structures
\cite{Tel,BD}.
We restrict ourselves to $\SL$-opers.

Let $E_3$ be a  $\SL$-bundle over a Riemann curve $\Si_{g,n}$ of genus $g$ with
$n$ marked points. It is a $\SL$-{\em oper} if
 there exists a flag filtration
$E_3\supset  E_2\supset   E_1\supset    E_0=0$ and a connection
 that acts as
$\nabla:~E_j\subset E_{j+1}\otimes\Om^{(1,0)}(\Si_{g,n})$.
Moreover, $\nabla$ induces an isomorphism
$E_j/E_{j-1}\to E_{j+1}/E_{j}\otimes\Om^{(1,0)}(\Si_{g,n})$.
We assume that the connection has poles in the marked points.
It possible to choose $E_1=\Om^{-1,0}(\Si_{g,n})$.
It means that locally the connection can represented as
\beq{6.1}
\nabla=\p -\thmat{0}{1}{0}{0}{0}{1}{W}{T}{0}.
\eq
This connection is equivalent to the third order differential operator
\beq{6.2}
\p^3-T\p-W:~\Om^{(-1,0)}(\Si_{g,n})\to\Om^{(2,0)}(\Si_{g,n}).
\eq
We assume that in  neighborhoods  of marked points $W$ and $T$
behave as
\beq{6.3}
T|_{z\rar x_a}\sim\frac{T^a_{-2}}{(z-x_a)^2}+\frac{T^a_{-1}}{(z-x_a)}+\ldots
\eq
\beq{6.4}
W|_{z\rar x_a}\sim
\frac{W^a_{-3}}{(z-x_a)^3}+\frac{W^a_{-2}}{(z-x_a)^2}+\frac{W^a_{-1}}{(z-x_a)}
+\ldots
\eq

Define the space $V_3$ of $\SL$-opers as the space of the third order
differential operators (\ref{6.2}) on $\Si_{g,n}$ with the coefficients $T$ and $W$
satisfying (\ref{6.3}),(\ref{6.4}).
\bigskip

{\bf 2. Lie algebroid over $\SL$-opers.}
Consider a vector bundle ${\cal A}_3$ over $V_3$. The space of sections
${\cal D}_2=\G({\cal A}_3)$ are the second order differential operators on
$\Si_{g,n}$ without constant terms. On a disk ${\cal A}_3$ can be trivialized
and the sections are represented as
$$
\ve^{(1)}=\ve^{(1)}(z,\bz)\frac{\p}{\p z},~~
\ve^{(2)}=\ve^{(2)}(z,\bz)\frac{\p^2}{\p z^2},
$$
$$
\ve^{(1)}\in {\cal D}^1,~~\ve^{(2)}\in {\cal D}^2,~~
{\cal D}_2={\cal D}^1\oplus{\cal D}^2.
$$
In addition, we assume that $\ve^{(1)},~\ve^{(2)}$
vanish holomorphically at the marked points as
\beq{6.5}
\ve^{(1)}\sim r^{(1)}_a(z-x_a)+o(z-x_a),~~
\ve^{(2)}\sim r^{(2)}_a(z-x_a)^2+o(z-x_a)^2.
\eq

We equip ${\cal A}_3$ with the structure of a Lie algebroid by defining the
Lie brackets on ${\cal D}_2$ and the anchor.
The second order differential operators do not generate
a closed algebra with respect to the standard commutators.
Moreover, they cannot be defined
invariantly on Riemann curves in contrast with the first order
differential operators.
We introduce a new brackets that goes around the both disadvantages.
The antisymmetric brackets on ${\cal D}_2$ are defined in the
following way.
\beq{6.6}
\lb\ve^{(1)}_1,\ve^{(1)}_2\rb=\ve^{(1)}_1\p\ve^{(1)}_2-\ve^{(1)}_2\p\ve^{(1)}_1.
\eq
\beq{6.7}
\lb\ve^{(1)},\ve^{(2)}\rb=\left\{
\begin{array}{cl}
-\ve^{(2)}\p^2\ve^{(1)}, &\in{\cal D}^1\\
-2\ve^{(2)}\p\ve^{(1)}+\ve^{(1)}\p\ve^{(2)},  &\in{\cal D}^2
\end{array}
\right.
\eq
\beq{6.8}
\lb\ve^{(2)}_1,\ve^{(2)}_2\rb=\left\{
\begin{array}{cl}
\frac{2}{3}[\p(\p^2-T)\ve^{(2)}_1]\ve^{(2)}_2 -
 \frac{2}{3}[\p(\p^2-T)\ve^{(2)}_2]\ve^{(2)}_1   , &\in{\cal D}^1\\
\ve^{(2)}_2\p^2\ve^{(2)}_1-\ve^{(2)}_1\p^2\ve^{(2)}_2,  &\in{\cal D}^2
\end{array}
\right.
\eq
The brackets (\ref{6.6}) are the standard Lie brackets of vector fields and therefore
${\cal D}^1$ is the Lie subalgebra of  ${\cal D}_2$.
The structure functions in (\ref{6.8}) depend on  the projective connection $T$.
Note that the brackets are consistent with the asymptotic (\ref{6.5}).

Now consider the bundle map ${\cal A}_3$ to $TV_3$ defined by the
anchor
\beq{6.9}
\de_{\ve^{(1)}}T=-2\p^3\ve^{(1)}+2T\p\ve^{(1)}+\p T\ve^{(1)},
\eq
\beq{6.10}
\de_{\ve^{(1)}}W=-\p^4\ve^{(1)}+3W\p\ve^{(1)}+\p W\ve^{(1)}+T\p^2\ve^{(1)},
\eq
\beq{6.11}
\de_{\ve^{(2)}}T=\p^4\ve^{(2)}-T\p^2\ve^{(2)}+(3W-2\p T)\p\ve^{(2)}+
(2\p W-\p^2T)\ve^{(2)},
\eq
\beq{6.12}
\de_{\ve^{(2)}}W=\frac{2}{3}\p^5\ve^{(2)}-\frac{4}{3}T\p^3\ve^{(2)}-
2\p T\p^2\ve^{(2)}+
\eq
$$
(\frac{2}{3}T^2-2\p^2T+2\p W)\p\ve^{(2)}+
(\p^2W-\frac{2}{3}\p^3T+\frac{2}{3}T\p T)\ve^{(2)}.
$$

\begin{theor}
The vector bundle ${\cal A}_3$ over the space of $\SL$-opers $V_3$ is a Lie
algebroid with the brackets (\ref{6.6}),(\ref{6.7}),(\ref{6.8}) and the anchor map
(\ref{6.9})-(\ref{6.12}).
\end{theor}
{\sl Proof}.
The algebroid structure follows from the identity
$$
[\de_{\ve^{(j)}_1},\de_{\ve^{(k)}_2}]=\de_{\lb\ve^{(j)}_1,\ve^{(k)}_2 \rb},~~
(j,k=1,2).
$$
The proof of this relation is straightforward, though is long and the calculations
were performed by the MAPLE. $\Box$

The SAJI (\ref{5.5}) in ${\cal A}_3$ takes the form
\beq{6.14}
\lb\lb\ve^{(2)}_1,\ve^{(2)}_2\rb,\ve^{(2)}_3\rb^{(1)}-
(\ve^{(2)}_1\p\ve^{(2)}_2-\ve^{(2)}_2\p\ve^{(2)}_1)\de_{\ve^{(2)}_3}T+{\rm c.p.}(1,2,3)
=0,
\eq
\beq{6.15}
\lb\lb\ve^{(2)}_1,\ve^{(2)}_2\rb,\ve^{(1)}_3\rb^{(1)}-
(\ve^{(2)}_1\p\ve^{(2)}_2-\ve^{(2)}_2\p\ve^{(2)}_1)\de_{\ve^{(1)}_3}T=0.
\eq
The brackets here correspond to the product of
structure functions in the left hand side
of (\ref{5.5}) and the superscript $(1)$ corresponds to the ${\cal D}^1$ component.
For the  rest  brackets the Jacobi identity is the standard one.
The origin of the brackets and the anchor representations follow from the
matrix description of $\SL$-opers (\ref{6.1}).
Consider the set $G_3$ of automorphisms of the bundle $E_3$
\beq{6.15a}
A\to f^{-1}\p f-f^{-1}Af
\eq
that preserve the $\SL$-oper structure
\beq{6.16}
f^{-1}\p f-
f^{-1}\thmat{0}{1}{0}{0}{0}{1}{W}{T}{0}f=
\thmat{0}{1}{0}{0}{0}{1}{W'}{T'}{0}.
\eq
It is clear that $G_3$ is the Lie groupoid over  $V_3=\{W,T\}$ with
$l(f)=(W,T)$, $~r(f)=(W',T')$, $~f\to <W,T|f|W',T'> $.
The left identity map is the $\SL$  subgroup of $G_3$
$$
P\exp(-\int^z_{z_0} A(W,T))\cdot C\cdot P\exp(\int^z_{z_0} A(W,T)),
$$
where $C$ is an arbitrary matrix from $\SL$ and $A(W,T))$ has the oper structure (\ref{6.1}).
The right identity map has the same
form with $(W,T)$ replaced by $(W',T')$.

The local version of (\ref{6.16}) takes the form
\beq{6.17}
\p X-\left[\thmat{0}{1}{0}{0}{0}{1}{W}{T}{0},X\right]=
\thmat{0}{0}{0}{0}{0}{0}{\de W}{\de T}{0}.
\eq
It is the sixth order linear differential system for the matrix elements of the
traceless matrix $X$. The matrix elements $x_{j,k}\in\Om^{(j-k,0)}(\Si_{g,n})$
 depend on two arbitrary fields $x_{23}=\ve^{(1)},~x_{13}=\ve^{(2)}$. The
solution takes the form
\beq{6.17b}
X=\thmat{x_{11}}{x_{12}}{\ve^{(2)}}{x_{21}}{x_{22}}
{\ve^{(1)}}{x_{31}}{x_{32}}{x_{33}},
\eq
$$
x_{11}=\frac{2}{3}(\p^2-T)\ve^{(2)}-\p\ve^{(1)},~
x_{12}=\ve^{(1)}-\p\ve^{(2)},
$$
$$
x_{21}=\frac{2}{3}\p(\p^2-T)\ve^{(2)}-\p^2\ve^{(1)}+W\ve^{(2)},~
x_{22}=-\frac{1}{3}(\p^2-T)\ve^{(2)},
$$
$$
x_{31}=\frac{2}{3}\p^2(\p^2-T)\ve^{(2)}-\p^3\ve^{(1)}+\p(W\ve^{(2)})+W\ve^{(1)},
$$
$$
x_{32}=\frac{1}{3}\p(\p^2-T)\ve^{(2)}-\p^2\ve^{(1)}+W\ve^{(2)}+T\ve^{(1)},
$$
$$
x_{33}=-\frac{1}{3}(\p^2-T)\ve^{(2)}+\p\ve^{(1)}.
$$
The matrix elements of the commutator $[X_1,X_2]_{13}$, $[X_1,X_2]_{23}$
give rise to the brackets (\ref{6.6}),\\(\ref{6.7}),(\ref{6.8}). Simultaneously,
from (\ref{6.17}) one obtain the anchor action (\ref{6.9})-(\ref{6.12}).

Consider the cohomology of ${\cal A}_3$.
There is a nontrivial cocycle corresponding to $H^1({\cal A}_3)$
with two components
\beq{6.17a}
c^{(1)}=\int_{\Si_{g,n}}\ve^{(1)}\bp T,
~~c^{(2)}=\int_{\Si_{g,n}}\ve^{(2)}\bp W.
\eq
It follows from the asymptotic of the sections (\ref{6.5}) and the fields
$T$ (\ref{6.3}), $W$ (\ref{6.4}) that the contributions
 from the marked points are equal
$$
c^{(1)}\rar \sum_{a=1}^n r^{(1)}_aT_{-2,a},~~
c^{(2)}\rar \sum_{a=1}^n r^{(2)}_aW_{-3,a}.
$$

The cocycle allows to shift the anchor action
$$
\hat{\de}_{\ve^{(j)}}f(W,T)=<\de_{\ve^{(j)}}W|\frac{\de f}{\de W}>+
<\de_{\ve^{(j)}}T|\frac{\de f}{\de T}>+c^{(j)},~~(j=1,2).
$$

There exists the $2g$ central extensions $c_\al$ of the algebra ${\cal A}_3$,
provided by the nontrivial cocycles from  $H^2({\cal A}_3,V_3)$. They are the
non-contractible contour integrals $\ga_\al$
\beq{6.16a}
c_\al(\ve^{(j)}_1,\ve^{(k)}_2)=\oint_{\ga_\al}\la(\ve^{(j)}_1,\ve^{(k)}_2),
~~(j,k=1,2),
\eq
where
$$
\la(\ve^{(1)}_1,\ve^{(1)}_2)=\ve_1^{(1)}\p^3\ve_2^{('1)},~~
\la(\ve^{(1)}_1,\ve^{(2)}_2)=\ve^{(1)}_1\p^4\ve^{(2)}_2,
$$
$$
\la(\ve^{(2)}_1,\ve^{(2)}_2)=
(\p^2-T)\ve_1^{(2)}\p(\p^2-T)\ve_2^{(2)}
+2(\p^2\ve_1^{(2)}\p(\p^2-T)\ve_2^{(2)}-\p^2\ve_2^{(2)}\p(\p^2-T)\ve_1^{(2)}).
$$
It can be proved that $sc^j=0$
(\ref{5.9}) and that $c^j$ is not exact. The proof is based on the matrix realization
of $\G({\cal A}_3)$ (\ref{6.17b}) and the two-cocycle (\ref{8.3}) of ${\cal A}_{\SL}$.
These cocycles allow to construct the central extensions of $\hat{\cal A}_3$:
$$
\lb(\ve_1^{(j)},\sum_\al k^{(j)}_\al),(\ve_2^{(m)},\sum_\al k^{(m)}_\al)\rb_{c.e.}=
(\lb\ve_1^{(j)},\ve_2^{(m)}\rb,\sum_\al c_\al(\ve_1^{(j)},\ve_2^{(m)})).
$$
\bigskip

{\bf 3. Hamiltonian algebroid over $W_3$-gravity}.
Let ${\cal R}_3$ be the affinization of the cotangent bundle $T^*V_3$
 to the space of $\SL$-opers $V_3$. The dual fields are the Beltrami differentials
$\mu$ and the differentials $\rho\in\Om^{(-2,1)}(\Si_{g,n})$.
We assume that   near the marked points $\rho$
 has the form
\beq{6.19}
\rho|_{z\rar x_a}\sim
(t^{(2)}_{a,0}+t^{(2)}_{a,1}(z-x^0_a))\bp\chi_a(z,\bz).
\eq
The space   ${\cal R}_3$ is the classical phase space for the $W_3$-gravity
\cite{P,GLM,BFK}. The symplectic form on ${\cal R}_3$ has the canonical form
\beq{6.20}
\om=\int_{\Si_{g,n}}\de T\wedge\de\mu+\de W\wedge\de\rho.
\eq

According to the general theory the anchor (\ref{6.9})-(\ref{6.12})
can be lifted from $V_3$ to ${\cal R}_3$.
This lift is nontrivial owing to the cocycle (\ref{6.17a}).
It follows from (\ref{5.16a}) that the anchor action on $\mu$ and
$\rho$ takes the form
\beq{6.21}
\de_{\ve^{(1)}}\mu=-\bar{\p}\ve^{(1)}-\mu\p\ve^{(1)}+\p\mu\ve^{(1)}-
\rho\p^2\ve^{(1)},
\eq
\beq{6.22}
\de_{\ve^{(1)}}\rho=-2\rho\p\ve^{(1)}+\p\rho\ve^{(1)},
\eq
\beq{6.23}
\de_{\ve^{(2)}}\mu=\p^2\mu\ve^{(2)}-\frac{2}{3}\left[(\p(\p^2-T)\rho)\ve^{(2)}
-(\p(\p^2-T)\ve^{(2)})\rho\right],
\eq
\beq{6.24}
\de_{\ve^{(2)}}\rho=-\bar{\p}\ve^{(2)}+(\rho\p^2\ve^{(2)}-\p^2\rho\ve^{(2)})
+2\p\mu\ve^{(2)}-\mu\p\ve^{(2)}.
\eq
There are two Hamiltonians, defining by the anchor and by the cocycle (see(\ref{5.17}))
$$
h^{(1)}=\int_{\Si_{g,n}}(\mu\de_{\ve^{(1)}}T+\rho\de_{\ve^{(1)}}W)+c^{(1)},~~
h^{(2)}=\int_{\Si_{g,n}}(\mu\de_{\ve^{(2)}}T+\rho\de_{\ve^{(2)}}W)+c^{(2)}.
$$
After the integration by part they take the form
$$
h^{(1)}=\int_{\Si_{g,n}}\ve^{(1)}F^{(1)},~~
h^{(2)}=\int_{\Si_{g,n}}\ve^{(2)}F^{(2)},
$$
where $F^{(1)}\in\Om^{(2,1)}(\Si_{g,n})$, $F^2\in\Om^{(3,1)}(\Si_{g,n})$
\beq{F1}
F^{(1)}=-\bp T-\p^4\rho+T\p^2\rho-(3W-2\p T)\p\rho-
\eq
$$
-(2\p W-\p^2 T)\rho+2\p^3\mu-2\p T\mu-\p T\mu,
$$
\beq{F2}
F^{(2)}=-\bp W-\frac{2}{5}\p^5\rho+\frac{4}{3}T\p^3\rho+2\p T\p^2\rho+
(-\frac{2}{3}T^2+2\p^2 T-2\p W)\p\rho+
\eq
$$
+(-\p^2W+\frac{2}{3}\p^3T-\frac{2}{3}T\p T)\rho+\p^4\mu
-3W\p\mu-\p W\mu-T\p^2\mu.
$$
They carry out the moment map
$$
m=(m^{(1)}=F^{(1)},m^{(2)}=F^{(2)}): {\cal R}_3\to \G^*({\cal A}_3).
$$
The elements of $\G^*({\cal A}_3)$ are singular at the marked points. In addition
to $y$ (\ref{8.13a}) there are $v$ dual to $\ve^{(2)}$ (\ref{6.5})
$$
v\sim c_{1,a}\p^2\de(x_a)+c_{2,a}\p^3\de(x_a)+\ldots.
$$
Let $m^{(1)}$ is defined as in (\ref{8.15}) and
$$
m^{(2)}=\sum_{a=1}^nW^a_{-3}\p^2\de(x_a).
$$
Then the coadjoint action of ${\cal D}_2$ preserve $m=(m^{(1)},m^{(2)})$.
The moduli space  ${\cal W}_3$ of the $W_3$-gravity
( $W_3$-geometry) is the symplectic quotient with respect to the groupoid
${\cal G}_3$ action
$$
{\cal W}_3={\cal R}_3//{\cal G}_3=\{F^1=m^{(1)},F^2=m^{(2)}\}/{\cal G}_3.
$$
It has dimension
$$
\dim{\cal W}_3=16(g-1)+6n.
$$
The term $6n$ comes from the coefficients
$(T_{-1}^a,W^a_{-1},W^a_{-2}),~a=1,\dots,n$ and the dual to them
$(t^{(1)}_{a,0},,t^{(2)}_{a,0},t^{(2)}_{a,1}),~a=1,\dots,n$.

The prequantization of ${\cal W}_3$ can be realized in the space of sections
of a linear bundle ${\cal L}$ over the space of orbits
 $\ti{V}_3\sim V_3/G_3$. The sections are functionals $\Psi(T,W)$ on $V_3$ satisfying
the following conditions
$$
\hat{\delta}_{\ve^{(j)}}\Psi(T,W):=
<\de_{\ve^{(j)}}W|\frac{\de \Psi}{\de W}>+
<\de_{\ve^{(j)}}T|\frac{\de \Psi}{\de T}>+c^{(j)}\Psi=0,~~(j=1,2).
$$
Presumably, the bundle ${\cal L}$ can be identified with the determinant bundle
 $\det(\p^3-T\p-W)$.

The moment equations $F^{(1)}=m^{(1)},~F^{(2)}=m^{(2)}$ are the consistency conditions
for the linear system
\beq{6.26a}
\left\{
\begin{array}{l}
(\p^3-T\p-W)\psi(z,\bz)=0,\\
(\bp +(\mu-\p\rho)\p +\rho\p^2
+\frac{2}{3}(\p^2-T)\rho-\p\mu)\psi(z,\bz)=0,
\end{array}
\right.
\eq
where $\psi(z,\bz)\in\Om^{-1,0}(\Si_{g,n})$. We will prove this statement below.
The last equation  represents the deformation of the antiholomorphic
operator $\bp$ (or more general $\bp+\mu\p$ as in (\ref{8.16}))
by the second order differential
operator $\p^2$. The left hand side is the exact form of the deformed operator
when it acts on $\Om^{-1,0}(\Si_{g,n})$. This deformation cannot
be supported by the structure of a Lie algebra  and one leaves with
the Hamiltonian algebroid symmetries.

Instead of the symplectic reduction one can apply the BRST construction.
Then cohomology of the moduli space  ${\cal W}_3$ are
isomorphic to
$H^j(Q)$. To construct the BRST complex we introduce the ghosts
fields $\eta^{(1)},\eta^{(2)}$ and their momenta ${\cal P}^{(1)},{\cal P}^{(2)}$.
Then it follows from Theorem 2.1 that for
$$
\Om=\sum_{j=1,2}h^{(j)}(\eta^{(j)})+
\oh\sum_{j,k,l=1,2}\int_{\Si_{g,n}}(\lb\eta^{(j)},\eta^{(k)}\rb{\cal P}^{(l)})
$$
the operator  $QF=\{F,\Om\}$ is nilpotent and define the BRST cohomology
in the complex
$$
\bigwedge{}^\bullet({\cal D}_2\oplus{\cal D}_2^*)\otimes C^\infty({\cal R}_3).
$$
\bigskip

{\bf 4. Chern-Simons derivation} \cite{BFK}.
Consider the Chern-Simons functional on $\Si_{g,n}\oplus {\bf R}^+$
$$
S=\int_{\Si_{g,n}\oplus {\bf R}}\tr({\cal A}d{\cal A}+\frac{2}{3}{\cal A}^3)+
\sum_{a=1}^n\int_{\bf R}\tr(p_ag_a^{-1}\p_tg_a),~~
({\cal A}=(A,\bA,A_t).
$$
 Introduce $n$ Wilson lines $W_a(A_t)$
along the time directions and located at the marked points
$$
W_a(A_t)=P\exp \tr(p_a\int A_t),~a=1,\ldots,n.
$$
In the hamiltonian picture the components
$A,\bA,\bfp,\bfg$ are elements of the phase space  with the symplectic form
(\ref{8.4a}) while $A_t$ is the Lagrange multiplier for the first class constraints
(\ref{8.5a}).

The phase space ${\cal R}_3$ can be derived from the phase space
 of the Chern-Simons. The flatness condition (\ref{8.5a}) generates
the gauge transformations
\beq{5.31}
A\to f^{-1}\p f+f^{-1}Af,~~\bA\to f^{-1}\bp f+f^{-1}\bA f,~~
p_a\to f_a^{-1}p_af_a,~~g_a\to g_af_a.
\eq
The result of the gauge fixing with respect to the whole gauge group $G_\SL$
is the moduli
space ${\cal M}^{flat}_3$ of the flat $\SL$ bundles over $\Si_{g,n}$.

Let $P$ be the maximal parabolic subgroup of $\SL$ of the form
$$
P=\thmat{*}{*}{0}{*}{*}{0}{*}{*}{*},
$$
and $G_P$ be the corresponding gauge group.
First we partly fix the gauge with respect to $G_P$.
A generic connection $\nabla$ can be gauge transformed by $f\in G_P$ to
  the form (\ref{6.1}).
It follows from (\ref{8.5a}) that $A$ has simple poles at the marked points.
To come to $V_3$
one should respect the behavior of the matrix elements at the marked points
(\ref{6.3}),(\ref{6.4}). For this purpose we use an additional singular gauge transform
by the diagonal matrix
$$
h=\prod_{a=1}^n\chi_a(z,\bz)\di(z-x_a,1,(z-x_a)^{-1}).
$$
The resulting gauge group we denote $G_{(P,h)}$.

The form of $\bA$ can be read off from (\ref{8.5a})
\beq{5.32}
\bA=\thmat{a_{11}}
{a_{12}}
{-\rho}
{a_{21}}
{a_{22}}
{-\mu}
{a_{31}}
{a_{32}}
{a_{33}}
\eq
$$
a_{11}=-\frac{2}{3}(\p^2-T)\rho+\p\mu,~~
a_{12}=-\mu+\p\rho,
$$
$$
a_{21}=-\frac{2}{3}\p(\p^2-T)\rho+\p^2\mu-W\rho,~~
a_{22}=\frac{1}{3}(\p^2-T)\rho,
$$
$$
a_{31}=-\frac{2}{3}\p^2(\p^2-T)\rho+\p^3\mu-\p(W\rho)-W\mu,
$$
$$
a_{32}=-\frac{1}{3}\p(\p^2-T)\rho+\p^2\mu-W\rho-T\mu,~~
a_{33}=\frac{1}{3}(\p^2-T)\rho-\p\mu.
$$
The flatness (\ref{8.5a})  for the special choice $A$ (\ref{6.1}) and
$\bA$ (\ref{5.32}) gives rise to
the moment constraints $F^{(2)}=0,~F^{(1)}=0$. Namely, one has
$F(A,\bA)|_{(3,1)}=F^{(2)}$ (\ref{F1}), $F(A,\bA)|_{(2,1)}=F^{(1)}$ (\ref{F2}),
while the other matrix
elements of $F(A,\bA)$ vanish identically.
At the same time, the matrix linear
system (\ref{8.5c}) coincides with  (\ref{6.26a}).
In this way, we come to the matrix description of the moduli space
${\cal W}_3$.

The cocycles $c_\al(\ve^{(j)}_1,\ve^{(k)}_2)$ (\ref{6.16a}) can be derived from
the two-cocycle (\ref{8.3}) of ${\cal A}_{\SL}$. Substituting in (\ref{8.3})
the matrix realization of $\G({\cal A}_3)$ (\ref{6.17b}), one come to (\ref{8.3}).

The groupoid action on $A,\bA$ plays the role of the rest gauge
transformations that complete the $G_P$ action to the $G_{\SL}$ action.
The algebroid symmetry arises in this theory as a result of the partial
gauge fixing by $G_{(P,h)}$. Thus we come to the following diagram.

\bigskip
$$
\begin{array}{rcccl}
           &\fbox{${\cal R}_{\SL}$}&                       &                    &\\
           &       |                      &\searrow{G_{(P,h)}}&           &\\
G_{\SL} &        |                     &                    &\fbox{${\cal R}_3$}&\\
           &\downarrow    &                   &\downarrow &\G({\cal A}^H_3)\\
           &\fbox{${\cal M}^{flat}_{\SL}$}&       &\fbox{${\cal W}_3$}&\\
\end{array}
$$
\bigskip

The tangent space to ${\cal M}^{flat}_{\SL}$ at the point $A=0,\bA=0,p_a=0,g_a=id$
coincides with the tangent space to ${\cal W}_3$ at the point
$W=0,T=0,\mu=0,\rho=0$. Their dimension is $16(g-1)+6n$. But their global
structure is different and the diagram cannot be closed by the horizontal
isomorphisms. The interrelations between ${\cal M}^{flat}_{\SL}$ and ${\cal W}_3$
were analysed in \cite{H2,Go}.

\section{Poisson sigma-model}
\setcounter{equation}{0}

The starting point in the description of the Poisson sigma-model
is a  manifold $M$, $\dim M=n$ endowed with
a Poisson bivector $\al^{jk}$ \cite{I,SS}.
It means that in local coordinates
$$
\{f(x),g(x)\}=
\al^{jk}\p_jf(x)\p_kg(x),~~(p_j=\p_{x^j},~x=(x_1,\ldots,x_n)\in M),
$$
and
$$
\al^{jk}(x)=-\al^{kj}(x), $$
\beq{7.1}
\p_i\al^{jk}(x)\al^{im}(x)+{\rm c.p.}(j,k,m)=0.
\eq
The manifold $M$ is the target space of the model. The space-time is
the unit disk $L=\{|z|\leq 1\}$.
There are two types of fields.
First one is a map
$$
X:~L\to M~~X(z,\bz)=(X^1,\ldots,X^n).
$$
Next, there is the one-form on $L$ taking values in the pull-back by
$X$ of the cotangent bundle $T^*M$:
$$
\xi(z,\bz)=(\xi_{1},\ldots,\xi_{n}),~~
\xi_{k}=\xi_{k,z}(z,\bz)dz+\xi_{k,\bz}(z,\bz)d\bz.
$$
The action is the functional
\beq{7.2}
S[X,\xi]=\int_L\xi_j(z,\bz)dX^j(z,\bz)+
\oh\al^{mn}(X)\xi_m(z,\bz)\xi_n(z,\bz).
\eq
\bigskip

{\bf 1. Hamiltonian description of the Poisson $\sigma$-model}.
Let $(t,\phi)$ be the polar coordinates on $L$ and $t$ play the role of time.
Then
$$
\xi_j=\xi_{j,t}dt+\xi_{j,\phi}d\phi,~~
dX^j=\p_tX^jdt+\p_\phi X^jd\phi.
$$
 In the hamiltonian form the phase space ${\cal R}$ of the model is the space of fields
$$
X(z,\bz)=(X^1,\ldots,X^n),~~\xi_\phi=(\xi_{\phi,1},\ldots,\xi_{\phi,n})
$$
on a circle endowed with the symplectic form
\beq{7.2a}
\om=<d\xi_{\phi}|d X>=\f1{2\pi}\int_{S^1}d\xi_{\phi,j}(z,\bz) dX^j(z,\bz).
\eq
In fact, the action (\ref{7.2}) takes the form
$$
S[X,\xi]=\int_L(\xi_{\phi,j}\p_tX^j)-(\xi_{t,j}F^j).
$$
Here $F^k=0$ are the $n$ first class constraints
\beq{7.3}
F^j:=\p_\phi X^j+\al^{jk}(X)\xi_{\phi,k}=0,
\eq
and $\xi_{j,t}$ are the Lagrange multipliers.
They generates the symmetries of $\om$
\beq{7.4}
\de_\ep X^k=\al^{kj}(X)\ep_j,
\eq
\beq{7.5}
\de_\ep\xi_m=-\p_\phi\ep_m+\p_m\al^{kj}\ep_k\xi_{\phi,j},
\eq
where $\ep_k$ are the sections of $X^*(T^*M)$.
\bigskip

{\bf 2. Hamiltonian Lie algebroid symmetries}.
Introduce the following brackets for the gauge transformations.
$$
\lb\ep,\ep'\rb_i=\p_i\al^{jk}(X)\ep_j\ep'_k.
$$
The structure functions  depend on $X$. Let $V$ be the space of smooth maps
$X:~S^1\to M$. Consider the bundle ${\cal A}$ over $V$ with sections
$\G({\cal A})=X^*(T^*M)=\{\ep\}$.
It is a Lie algebroid with the anchor map (\ref{7.4}).
In fact, it is straightforward to derive from  (\ref{7.4}) and (\ref{7.5})
that $\lb\de_\ep,\de_{\ep'}\rb=\de_{\lb\ep,\ep'\rb}$.

Consider the one-cocycle
\beq{7.7}
c(X,\ve)=\f1{2\pi}\int_{S^1}\ep_j\p_\phi X^j.
\eq
It is easy to see that it represents an element of $H^1({\cal A},V)$.
Therefore, one can extend the anchor action (\ref{7.4})
$$
\hat{\de}_\ep f(X)={\de}_\ep f(X)+c(X,\ve).
$$

As it follows from (\ref{7.5}), the phase space ${\cal R}$ is
the affine bundle over the cotangent bundle $T^*V$.
 The gauge transformations
(\ref{7.5}) is the lift of the anchor (\ref{7.4}) by means of the cocycle (\ref{7.7}).
Moreover, according to Lemma 2.1 the Hamiltonians, defined by the constraints
(\ref{7.3})
$$
h_{\ep_j}=\f1{2\pi}\int\ep_jF^j,
$$
(no summation on $j$) give rise to the Hamiltonian algebroid ${\cal A}^H$
over ${\cal R}$.

Following our approach we interpret the constraints  (\ref{7.3}) as consistency
conditions for a linear system. First, introduce the operator $B$ from
$X^*(T^*M)$ to $X^*(TM)$ and the corresponding linear system
\beq{7.4a}
B^{jm}(X)\psi_m=0,~~B^{jm}(X)=\la+\al^{jm}(X)
\eq
where $\psi_m$ is a section of $X^*(T^*M)$ . The second equation is
determined by the operator $A:X^*(T^*M)\to X^*(T^*M)$
\beq{7.5a}
A_m^k\psi_k=0,~~A_m^k=-\p_\phi+\p_m\al^{ks}\xi_{\phi,s}.
\eq
\begin{lem}
Let the Poisson bivector satisfies the non-degeneracy condition:\\
 the matrix $a_i^j=(\p_i\al^j)^m$ is non-degenerate
on $V$ for some $m$.\\
 Then the constraints  (\ref{7.3}) are the consistency
conditions for (\ref{7.4a}) and (\ref{7.5a}).
\end{lem}
{\sl Proof.}
Define the dual operator
$$
A^*:X^*(TM)\to X^*(TM),~~(A^*)_i^j=(-\p_\phi-\p_i\al^{js}\xi_{\phi,s}),
$$
It gives rise to the equation
\beq{7.6a}
(A^*)_i^j\psi^i=0.
\eq
The consistency condition of these equations is the operator equation
$$
BA-A^*B=0.
$$
After substitution in it the expressions for $A,A^*,B$ and applying the Jacobi identity
(\ref{7.1}) one comes to the equality
$$
(\p_\phi X^i+\al^{is}\xi_s)\p_i\al^{jm}\psi_m=0.
$$
The later is equivalent to the constraint equation (\ref{7.1}) if $\al$ is
non-degenerate in the above sense. $\Box$

Let $\Psi(X)$ be a smooth functional on $V$ satisfying the following condition
\beq{LB}
\hat{\de}_{\ep_j}\Psi:=\f1{2\pi}\int_{S^1}\ep_j
\al^{kj}(X)\frac{\de}{\de X^k}\Psi(X)+
\left(\f1{2\pi}\int_{S^1}\ep_j\p_\phi X^j
\right)\Psi(X)=0.
\eq
Let $G$ be the Lie groupoid corresponding to the Lie algebroid ${\cal A}$.
Consider the space of orbits $\ti{V}=V/G$ and a line bundle ${\cal L}(\ti{V})$ over
$\ti{V}$
that has the space of sections $\G({\cal L})$ the functionals
$\Psi(X)$ on $V$ (\ref{LB}). Presumably, it is the determinant bundle
$\det(\la+\al^{jm}(X))$ coming from (\ref{7.4a}).
Consider the symplectic quotient
$$
{\cal R}^{red}={\cal R}//G^H=\{F^j=0\}/G^H,
$$
where $G^H$ is the Hamiltonian groupoid.
As it follows from the general construction $\G({\cal L})$ serves as the Hilbert
space in the prequantization of the phase space ${\cal R}^{red}$.

The quantization of ${\cal R}^{red}$ can be performed by the BRST technique.
The classical BRST complex is the set of fields
$$
\bigwedge{}^\bullet\left
(\G(X^*(TM))
\oplus\G(X^*(T^*M))
\right)\otimes C^\infty({\cal R}).
$$

Theorem 2.1 states that the BRST operator has rank one
$$
\Om=\f1{2\pi}\int_{S^1}\eta_jF^j+
\f1{\pi}\int_{S^1}\p_i\al^{kj}(X)\eta_j\wedge\eta_k{\cal P}^i,
$$
where $\eta=(\eta_1,\ldots\eta_n)$ are  dual to the gauge
generators $\ep$ and ${\cal P}$ are their momenta.
It means that the deformation of the Poisson bivector on $M$
does not affect the Lie algebraic form of $\Om$. This form of $\Om$
was found in \cite {SS}.

\section{Concluding Remarks}
 Let summarize the results and discuss some open problems.\\
(i) We defined the Hamiltonian algebroids. These objects arise in a natural
way in the hamiltonian systems with the first class constraints \cite{Ba}.
The BRST operator for these systems has an arbitrary rank and can be constructed
by the perturbation theory \cite{HT}.
On the other hand, it was suggested in \cite{RW} that another type of algebroids
- Courant algebroids - are related to the same systems of classical mechanics.
It would be interesting to establish relations between the Hamiltonian and
Courant algebroids.\\
(ii) The special kind of the Hamiltonian algebroids are defined over principle
affine space over
cotangent bundles. Any Lie algebroid defined over the base of the cotangent
bundles can be lifted to these Hamiltonian algebroids. The lifts are classified
by the first cohomology of the Lie algebroid.
The Hamiltonian algebroids  of this type are most closed to the Lie algebras
of Hamiltonian vector fields and has the same form of the BRST operator.
\\
(iii) The Lie algebroid over the space of $\SL$-opers on a Riemann curve with
marked points has the space of second order differential operators as
the space of  sections. It contains the Lie subalgebra of the first order
differential operators. After change the behavior of their coefficients
at the marked points this subalgebra just coincides with the Krichever-Novikov
algebra \cite{KN}. It will be interesting to lift this correspondence to the
higher order differential operators. Another open question is the structure of
opers and Lie algebroids defined on Riemann curves with double marked points.\\
(iv) Though the generalization
to $W_N, ~N>3$ is straightforward the limit $N\to\infty$, where the structure
of the strongly homotopy Lie algebras
 should be recovered, is looked obscure in our approach. \\
(v) The Chern-Simons derivation of the Hamiltonian algebroid in $W_3$-gravity
explain the origin of the algebroid symmetry as a result of the two step gauge
fixing. It will be plausible to have the same universal construction for
the Poisson sigma-model that responsible for the deformation quantization
of the Poisson brackets.

\setcounter{equation}{0}

\small{

}
\end{document}